\documentclass[a4paper,11pt]{article}
\pdfoutput=1 

\usepackage{jheppub} 

\usepackage[T1]{fontenc} 
\usepackage{caption}
\usepackage{comment}
\usepackage{amsmath}
\usepackage{graphicx}
\usepackage{tikz}
\usetikzlibrary{arrows,decorations.pathmorphing}
\usepackage[compat=1.1.0]{tikz-feynman}
\usepackage{amssymb}
\usepackage{bm,commath,mathtools,wasysym,xfrac}
\usepackage{pdfpages}
\usepackage{tcolorbox}
\usepackage{braket}
\usepackage{MnSymbol}
\usepackage{cancel}
\usepackage[english]{babel} 
\usepackage{blindtext}
\usepackage[normalem]{ulem}

\usepackage{cancel}
\usepackage{color}

\tikzset{
  graviton/.style={decorate, decoration={snake, amplitude=2pt, 
    segment length=6pt}, thick},
  source/.style={circle, fill=black, inner sep=2pt},
  vertex/.style={circle, draw=black, thick, fill=white, inner sep=2pt},
  static/.style={dashed, thick},
  vertexstat/.style={circle, draw=black, fill=magenta, inner sep=2pt},
  momentum/.style={->, >=stealth, thin}
}

\graphicspath{{./figs/}}

\newcommand{\be}{\begin{eqnarray}}
\newcommand{\ee}{\end{eqnarray}}
\newcommand{\bdm}{\begin{displaymath}}
\newcommand{\edm}{\end{displaymath}}
\newcommand{\bea}{\begin{align}}
\newcommand{\eea}{\end{align}}
\newcommand{\ba}{\begin{array}}
\newcommand{\ea}{\end{array}}
\newcommand{\pa}[1]{\left(#1\right)}
\newcommand{\paq}[1]{\left[#1\right]}
\newcommand{\pag}[1]{\left\{#1\right\}}
\newcommand{\ds}{\displaystyle}

\newcommand{\sgn}{\mathrm{sgn}}

\title{\boldmath 
Classical soft theorems meet the post-Newtonian expansion}

\author[]{Samim Akhtar}
\author[]{and Riccardo Sturani}

   \affiliation[]{ICTP South American Institute for Fundamental Research, \\
  Rua Dr. Bento Teobaldo Ferraz 271-2, 01140-070, São Paulo, SP, Brazil} 
  \affiliation[]{Instituto de Física Teórica UNESP - Universidade Estadual Paulista, \\
  Rua Dr. Bento Teobaldo Ferraz 271, 01140-070, São Paulo, SP, Brazil} 

\emailAdd{samim.akhtar@ictp-saifr.org}
\emailAdd{riccardo.sturani@unesp.br}

\abstract{Classical soft theorems govern the non-analytic structure of electromagnetic and gravitational radiation in the low-frequency regime, with the gravitational memory effect providing the canonical realization of the classical soft graviton theorem. While the soft behavior of radiation from hyperbolic scattering is well understood, its realization in gravitationally bound systems remains less explored.

In this work, we study the soft structure of gravitational waveforms generated by compact binaries, focusing on the scattering of emitted radiation off the static background curvature within the multipolar post-Newtonian framework. We isolate the infrared (IR) logarithms associated with hereditary effects, including the tail-of-memory, and their higher generalizations with $n$-tail insertions. The resulting amplitudes exhibit a Laurent expansion in frequency space consistent with the Sahoo–Sen logarithmic soft theorems, thereby establishing a direct correspondence between gravitational memory effects, their higher-order tail corrections, and the soft expansion. We also comment on ultraviolet (UV) logarithms arising from short-distance sensitivity of the multipole expansion and their associated renormalization-group running.}

\makeatletter
\gdef\@fpheader{}
\makeatother

\begin{document} 
	\maketitle
	\flushbottom

\section{Introduction}
\label{sec:intro}

The direct detection of gravitational waves (GW)s from compact binary coalescences~\cite{LIGOScientific:2016aoc,LIGOScientific:2026wfs}
has transformed GW astronomy into a precision science. As the sensitivity
of the current LIGO--Virgo--KAGRA network continues to improve \cite{LIGOScientific:2014pky,VIRGO:2026okv}, and with the third-generation
ground-based detectors (Einstein Telescope \cite{ET:2025xjr}, Cosmic Explorer \cite{Reitze:2019iox}), together with the space-borne LISA \cite{LISA:2024hlh}
mission on the horizon, increasingly accurate theoretical waveform models are essential \cite{Purrer:2019jcp,Thompson:2025hhc}.
In one of the most developed analytic frameworks to study the gravitational two-body problem, the post-Newtonian (PN) \cite{Blanchet:2013haa} approximation to General Relativity (GR),
beyond the leading quadrupole formula, the GWform receives a rich set
of \emph{hereditary} corrections: contributions that depend not only on the instantaneous
state of the source but on its entire past history~\cite{Blanchet:1987wq,Blanchet:1992br,Christodoulou:1991cr}.
 
Two partially overlapping, largely complementary frameworks have proven especially powerful for computing these
corrections. On one hand, the PN approximation has been
recast into an effective field theory, via
the Non-Relativistic General Relativity (NRGR) approach \cite{Goldberger:2004jt} and extended to spinning bodies \cite{Porto:2005ac,Porto:2016pyg}, and it organises GW emission
as a perturbative expansion with Feynman diagrams encoding the various
physical processes perturbatively in $v/c$, with $v^2\sim GM/r$, where $v$ is the relative velocity of binary constituents, $M$ their total mass and $r$ their separation. 
Higher-order PN conservative dynamics and hereditary effects within this framework have been systematically developed in~\cite{Foffa:2011ub,Foffa:2019yfl,Foffa:2019eeb,Foffa:2021pkg,Almeida:2021xwn,Almeida:2023yia,Almeida:2026clf}, with remarkable results as the leading logarithmic behaviour (or ``anomalous dimensions'') of radiative multipoles \cite{Goldberger:2009qd,Almeida:2021jyt,Fucito:2024wlg,Ivanov:2025ozg}, logarithmic contributions to the gauge invariant energy of circular orbits \cite{Blanchet:2019rjs}, and memory contributions to the waveform and effective action ~\cite{Blanchet:1997ji,Almeida:2024lbv,Porto:2024cwd}.

On the other hand, the \emph{post-Minkowskian} (PM) programme~\cite{Damour:2016gwp,Damour:2019lcq}
organises the dynamics as a perturbation in Newton's constant $G$ without
assuming small velocities; a remarkable renaissance in this direction was triggered by
the application of modern amplitudes technology~\cite{Cheung:2018wkq,Bern:2019nnu,Bern:2019crd}
including the double-copy and generalised unitarity methods \cite{Bern:2021dqo,Bern:2021yeh}, as well as
the eikonal approach developed in~\cite{Bjerrum-Bohr:2018xdl,Bjerrum-Bohr:2021vuf}, and the eikonal
operator methods ~\cite{DiVecchia:2021bdo,DiVecchia:2021ndb}, which are particularly suited for describing \emph{unbound} scattering processes. Classical observables from
amplitudes have also been extracted systematically via the KMOC
formalism~\cite{Kosower:2018adc}, and spinning degrees of freedom have been incorporated in~\cite{Guevara:2018wpp,Guevara:2019fsj,Maybee:2019jus}
connecting minimal-coupling amplitudes to the Kerr black hole. The boundary-to-bound
(B2B) correspondence~\cite{Kalin:2019rwq,Kalin:2019inp,Cho:2021arx} provides a systematic
path from PM-derived scattering results to bound-orbit observables.
 
The principal hereditary effects computed within the PN framework are well
established. The \emph{tail effect}~\cite{Blanchet:1987wq,Blanchet:1992br}
arises from the backscattering of emitted radiation off the static background curvature
generated by the total mass $M$. The \emph{GW memory
effect}~\cite{Zeldovich:1974gvh,Braginsky:1987kwo,Wiseman:1991ss}
is a permanent non-oscillatory \emph{displacement} of test masses; at \emph{non-linear} order it is
sourced by the stress-energy of the emitted radiation
itself~\cite{Christodoulou:1991cr,Thorne:1992sdb}. The \emph{tail-of-memory}~\cite{Trestini:2023wwg}
is their combination: the non-linear memory itself undergoes tail scattering. In the PN
framework, these contributions affect the radiative multipole at relative 1.5, 2.5, and
4\,PN order respectively~\cite{Blanchet:2013haa}, with tail-of-tail and higher interactions at 3\,PN
and beyond~\cite{Blanchet:1997jj}. 
 
However, the universality of gravitational radiation at low frequencies is encoded in non-perturbative \emph{soft theorems}. Weinberg's leading soft graviton
theorem~\cite{Weinberg:1965nx} produces the characteristic $1/\omega$ pole in the scattering amplitude that,
in the classical limit, corresponds to the \emph{displacement} memory effect. This connection
was established in a remarkable series of papers~\cite{He:2014laa,Strominger:2014pwa,Strominger:2016wns}, where it has been shown that the
\emph{memory effect}, the \emph{soft graviton theorem}, and the \emph{asymptotic BMS supertranslation
symmetry}~\cite{Bondi:1962px,Sachs:1962wk} are in connection to one another to form a so-called \emph{infrared triangle}.

The extension of these relations
to sub-leading order, connecting the \emph{spin} memory effect to the Cachazo--Strominger soft theorem \cite{Cachazo:2014fwa} and
superrotation symmetry, was developed in~\cite{Campiglia:2014yka,Campiglia:2015yka}. Spin memory is not a displacement-memory: it is the change in a
retarded-time integral of a magnetic type portion of the gravitational shear, which can be measured as a relative time delay of
counter-orbiting light rays.  Its source is angular-momentum flux rather
than energy flux~\cite{Pasterski:2015tva,Nichols:2017rqr} and it has a clear counterpart in soft-theorem. We review this distinct
observable in section~\ref{sec:soft_review}, but do not include it in the electric-parity
tail-of-memory ladder computed below.

By soft terms we mean terms non-analytic in $\omega$, with the property that either themselves or a finite order derivative of them are divergent for $\omega\to 0$.
Beyond the leading $1/\omega$ term, the soft expansion in four dimensions receives
\emph{logarithmic} corrections. The universality of the classical waveform dictates
that the time Fourier transform of the $1/r$ asymptotic perturbation
admits~\cite{Sahoo:2018lxl,Saha:2019tub,Sen:2024qzb}
\begin{align}
  \tilde{h}_{\mu\nu}(\omega, \hat{n})
  \overset{\omega\to 0}{=} & \omega^{-1} \, \tilde{h}_{\mu\nu}^{(0)}(\hat{n}) \, 
  + \, \log{\omega} \, \tilde{h}_{\mu\nu}^{(\log)}(\hat{n}) \, + \, \omega(\log{\omega})^2 \, \tilde{h}_{\mu\nu}^{(\log)^2}(\hat{n}) 
  + \cdots,
  \label{eq:soft_exp}
\end{align}
where $\hat{n}$ is a unit vector on the celestial sphere at null infinity. The coefficients $\tilde{h}_{\mu\nu}^{(0)}$ (displacement memory), $\tilde{h}_{\mu\nu}^{(\log)}$ (tail of memory), $\tilde{h}_{\mu\nu}^{(\log)^2}$ (tail-of-tail of memory) are \emph{universal}: they depend only on the asymptotic four-momenta of
the incoming and outgoing objects; they are independent of the details of the
interaction~\cite{Sahoo:2018lxl,Saha:2019tub,Sen:2024qzb}. This universality is a direct consequence of general coordinate invariance — the soft graviton theorem follows from the Ward identity of diffeomorphism symmetry acting on the asymptotic states, and is therefore insensitive to the specific form of the interactions in the bulk. As a result, the soft coefficients $\tilde{h}_{\mu\nu}^{(0)}$, $\tilde{h}_{\mu\nu}^{(\log)}$, and so on remain unchanged even in theories beyond GR, as long as the graviton remains massless and diffeomorphism invariance is preserved~\cite{Weinberg:1965nx,Sen:2024qzb}. The existence and structure
of these logarithmic terms was uncovered in a series of papers by Laddha and
Sen~\cite{Laddha:2018myi,Laddha:2018vbn,Laddha:2019yaj}, and the classical
logarithmic soft graviton theorem was proved and extended in a series of papers by
Sahoo and Sen~\cite{Sahoo:2018lxl,Saha:2019tub,Sahoo:2021ctw,Sahoo:2020ryf}. In particular, in ~\cite{Sen:2024qzb} Sen derives an all-order prediction for the leading-logarithm tower $\omega^{n-1}(\ln\omega)^n$ and finds that, for a binary black-hole merger
ending in a single massive remnant and massless radiation, the first two logarithmic soft contributions determined by the outgoing massive state vanish. 

Within this context, we highlight that terminology requires care.  In the PN-MPM literature on radiation from binary systems, a tail
refers solely to the \emph{drag} that slows the outgoing radiation and produces a phase shift in the
radiative field.  The soft-theorem ``tail to the memory'', by contrast, is the sum of this same
drag contribution and the contribution due to the logarithmic growth of the angular momentum of
the scattering bodies~\cite{Geiller:2024ryw}. Accordingly, the 3-graviton interaction involving 3 radiative ones studied in this work isolates the common radiative-drag sector, whereas the latter contribution is not relevant for the gravitationally bound systems we are considering here. In PN--MPM, the 4PN cubic hereditary interaction $M\times {\mathcal M}_{ij}\times {\mathcal M}_{kl}$ (being ${\mathcal M}_{ij}$ the \emph{canonical} mass quadrupole moment)
contains this drag sector together with pieces
irrelevant to the leading soft logarithm; these affect finite terms or subleading soft behavior. Observational signatures of these logarithmic terms were explored
in~\cite{Laddha:2018vbn}, and sub-subleading extensions
in~\cite{Sahoo:2020ryf}.

Complementary perspectives on the logarithmic hierarchy have been developed from the PM
side. In \cite{Alessio:2024onn} the soft energy
emission spectrum and the leading-logarithm tower have been computed in the PM expansion using the
classical soft factors, establishing the conjecture $\omega^{n-1}(\ln\omega)^n$ behaviour at all
orders in a PM context. The logarithmic soft theorem has also been placed in the context
of asymptotic symmetries — specifically the superrotation Ward identity — in~\cite{Campiglia:2019wxe,Donnay:2020lur,Choi:2024ajz}.

A subtlety that arises when 
comparing PM and PN results is the choice of BMS frame: amplitude-based results are more naturally expressed in the \emph{canonical} BMS frame where the asymptotic shear vanishes in the far past, while PN-multipolar PM (MPM) expanded formulas hold in the \emph{intrinsic} BMS frame, which retains the time-independent shear sourced by the Coulomb (linearized-Schwarzschild) fields of bodies in asymptotic uniform motion~\cite{Veneziano:2022zwh,Georgoudis:2023eke}. This distinction affects the asymptotic charges and the identification of the memory contribution in the waveform, and must be accounted for when comparing PN and PM results for the soft coefficients~\cite{Campiglia:2019wxe,Donnay:2020lur,Georgoudis:2023eke,Agrawal:2023zea,Compere:2019gft,Compere:2025tzr}. One of the present authors showed
in~\cite{Akhtar:2024lkk} that in the infinite-impact-parameter limit the entire tower, $\omega^n \ln{\omega}$,
can be derived using (sub)$^n$-leading soft graviton theorems via the KMOC
formalism~\cite{Kosower:2018adc}. Soft constraints on PM waveforms using the KMOC
approach were studied in~\cite{Bautista:2021llr}, and PM
waveforms from amplitudes have been computed in~\cite{Bautista:2019tdr,Manu:2020zxl,Bautista:2021inx}.
 
In contrast to the hyperbolic scattering case, the realisation of the soft-theorem
hierarchy in \emph{gravitationally bound} systems presents additional subtleties. In hyperbolic scattering, the asymptotic trajectories of the massive hard particles acquire logarithmic deviations at late times due to the long-range gravitational interaction, generating additional logarithmic contributions to the soft waveform beyond those from radiation backscattering. In a bound system, the particles never reach free asymptotic states — they remain bound for all time — so these trajectory-based contributions are absent. The explicit IR logarithms in the far-zone amplitudes therefore arise solely from the backscattering (tail) sector — the emitted radiation propagating to null infinity and scattering off the background Newtonian potential of the total mass $M$. Despite the extensive activity connecting PN hereditary effects
and soft theorems from the PM/amplitude
side~\cite{Bautista:2019tdr,Bautista:2021inx,Bern:2021yeh,Alessio:2024onn}, the
direct correspondence between the PN hereditary amplitude hierarchy — as developed
in NRGR ~\cite{Foffa:2019eeb,Almeida:2021xwn,Almeida:2021jyt} —
and the Sahoo--Sen logarithmic expansion has not been made explicit for the gravitationally bound binaries.

In this work we bridge that gap. Working within the NRGR framework in
$d = 3-\epsilon$ spatial dimensions with retarded propagators and renormalisation scale
$\tilde\mu^2 = \pi\mu^2 e^{-\gamma}$, we compute the one- and multi-loop emission
amplitudes corresponding to the non-linear memory, the tail-of-memory, and their
generalisations obtained by inserting an arbitrary number $n$ of static mass
insertions. Our main results are:
 
\begin{itemize}
 
\item The memory amplitude at 2.5\,PN is a one-loop diagram with no static
  propagator. Its TT radiative component has no IR divergence and yields only the Weinberg soft pole
  $\sim 1/\omega$, providing the PN realisation of the leading coefficient $\tilde{h}_{\mu\nu}^{(0)}$
  in~\eqref{eq:soft_exp}.
 
\item The tail-of-memory amplitude at 4\,PN (two-loop, one static propagator) produces
  an IR pole $1/\epsilon_{\rm IR}$ and the physical $\ln\omega$, reproducing the
  PN realization of the Sahoo--Sen coefficient $\tilde{h}_{\mu\nu}^{(\log)}$ in~\eqref{eq:soft_exp}.
 
\item The generalisation to $n$ mass insertions — the $(n)$-tail-of-memory at
  $(2.5 + 1.5n)$\,PN — yields the universal soft behaviour
  \begin{equation}
    \mathcal{A}_n\big|_{\omega\to 0}
    \;\sim\; (G M)^n \cdot E_{\rm rad} \cdot \omega^{n-1}(\ln\omega)^n,
  \end{equation}
  where $E_{\rm rad}$ denotes the total radiated energy. This provides the first PN/NRGR derivation of the all-order leading-logarithm tower conjectured in~\cite{Sen:2024qzb,Alessio:2024onn}.
 
\item The tower exponentiates into a universal soft-dressed factor,
  confirming the IR exponentiation structure of the radiation
  EFT~\cite{Porto:2012as,Almeida:2021xwn,Ivanov:2025ozg}. A related computation of the tail hierarchy via generalized unitarity appears in~\cite{Edison:2022cdu}, though in the self-energy context where the graviton is integrated out, rather than kept on-shell as an emitted radiation mode.
 
\item UV logarithms first arise at $n\ge 2$ and hence are subleading relative to the leading IR tower.
 
\end{itemize}
 
The paper is organised as follows. In section~\ref{sec:soft_review} we review the
classical soft theorems and their realisation in the PN--MPM framework, derivation of displacement memory and tail-of-memory from first principles, and the
separate spin-memory observable and its subleading-soft interpretation. In
section~\ref{sec:NRGR_memory} we compute these amplitudes in NRGR and analyse their
divergence structure. In section~\ref{sec:higher} we generalise to $n$ tail insertions.
In section~\ref{sec:IR_softexp} we discuss IR exponentiation, the Sahoo--Sen
correspondence, and the UV subtraction procedure. We close with conclusions in
section~\ref{sec:conclusions}. The memory kernel integrals, NRGR framework, master integrals, and
polarisation-sector analysis are in appendices~\ref{sec:app_memory_angular}, \ref{sec:app_nrgr}, \ref{sec:app_master}, and \ref{sec:app_polarsector} respectively.

\section{A brief review of classical soft theorems} \label{sec:soft_review}

Computing the GWform at a finite retarded time is a
challenging problem that requires detailed information about the full scattering or collision process. However, the behavior of the waveform at asymptotically early and late retarded times—i.e. long before and long after the main burst of radiation reaches the
detector—admits universal expressions. These depend solely on the momenta and spin of the incoming and outgoing objects, without any reference to the detailed dynamics
during the interaction. Such asymptotic features of the waveform are governed by the low-frequency behavior of its time Fourier transform, which in turn is determined by classical soft
theorems \cite{Sen:2024qzb}: 
\be
\bar{h}_{\mu\nu}(u,\hat{n}) = \begin{cases} A_{\mu\nu}^{+}(\hat{n}) + B_{\mu\nu}(\hat{n}) \, u^{-1} + F_{\mu\nu}(\hat{n}) \, u^{-2} \ln{u} + \mathcal{O}(u^{-2}) + \mathcal{O}(r^{-2}) \,, ~~~ \text{for} \, u \rightarrow +\infty \,,\\
A_{\mu\nu}^{-}(\hat{n}) + C_{\mu\nu}(\hat{n}) \, u^{-1} + G_{\mu\nu}(\hat{n}) \, u^{-2} \ln\pa{-u} + \mathcal{O}(u^{-2}) + \mathcal{O}(r^{-2}) \,, ~~~ \text{for} \, u \rightarrow -\infty \,,
\end{cases}
\ee
where $\bar{h}_{\mu\nu} = h_{\mu\nu} - \frac{1}{2}\eta^{\rho\sigma}h_{\rho\sigma} \eta_{\mu\nu}$.
In terms of the time Fourier transform, defined as
\be
\tilde{h}_{\mu\nu}(\omega,\hat n) = \int du \, e^{i \omega u} \, \bar{h}_{\mu\nu}(u,\hat n)\,,
\ee
then for small $\omega$, $\tilde{h}_{\mu\nu}(\omega)$ has an expansion of the form:
\begin{align}
\tilde{h}_{\mu\nu}(\omega, \hat{n})
  \overset{\omega\to 0}{=} & \omega^{-1} \, \tilde{h}_{\mu\nu}^{(0)}(\hat{n}) \, 
  + \, \log{\omega} \, \tilde{h}_{\mu\nu}^{(\log)}(\hat{n}) \, + \, \omega(\log{\omega})^2 \, \tilde{h}_{\mu\nu}^{(\log)^2}(\hat{n}) 
  + \cdots \,,
\end{align}
where the first term is the displacement memory. Here,
\begin{align}
    \tilde{h}_{\mu\nu}^{(0)} &= i ( A_{\mu\nu}^{+} - A_{\mu\nu}^{-}) \,, \cr
    \tilde{h}_{\mu\nu}^{(\log)} &= - (B_{\mu\nu} - C_{\mu\nu}) \,, \cr
    \tilde{h}_{\mu\nu}^{(\log)^2} &= \frac{i}{2} (F_{\mu\nu} - G_{\mu\nu})\,.
\end{align}
The leading coefficients $A_{\mu\nu}^{\pm}$ depend on the BMS frame, but their difference is insensitive to this frame choice. Working in the \emph{canonical} frame where $A_{\mu\nu}^{-} = 0$, we set $A_{\mu\nu}^{+} = A_{\mu\nu}$.
For 2-2 scattering, up to $\ln{\omega}$ terms they are given by
\be
A_{\mu\nu} = \frac{2 G}{R c^3 } \left(S^{(0)}_{\mu\nu}(\{p_a\},\hat{n}) - S^{(0)}_{\mu\nu}(\{p_a^\prime\},\hat{n})\right) \,,
\ee
and the tail-of-memory
\begin{align} \label{eq:sen_tom}
B_{\mu\nu} =& \frac{4 G^2}{R c^7} \Bigg(\sum_{a,b=1 | b \neq a}^2 S^{(1)}_{\mu\nu}(\{p_a^\prime \},\hat{n}) + \sum_{b=1}^2 (p_b^\prime \cdot \hat{n})S^{(0)}_{\mu\nu}(\{p_a^\prime\},\hat{n}) - \sum_{b=1}^2 (p_b \cdot \hat{n}) S^{(0)}_{\mu\nu}(\{p_a\},\hat{n}) \Bigg) \,, \cr
C_{\mu\nu} =& -\frac{4 G^2}{R c^7 } \Bigg(\sum_{a,b=1 | b \neq a}^2 S^{(1)}_{\mu\nu}(\{p_a \},\hat{n}) \Bigg) \,.
\end{align}
Here
\begin{align} 
S^{(0),\mu\nu}(\{p_a\},\hat{n}) &=\sum_{a=1}^2 \frac{p_a^{\mu}p_a^{\nu}}{p_a \cdot \hat{n}} \,, \cr
     S^{(1),\mu\nu}(\{p_a\},\hat{n}) &= (p_1 \cdot p_2)  \frac{(2(p_1 \cdot p_2)^2 - 3m_1^2 m_2^2)}{[(p_1 \cdot p_2)^2 -m_1^2 m_2^2]^{3/2}} \frac{\hat{n}_\rho }{p_a \cdot \hat{n}} p_a^{(\mu}\left( p_a \wedge p_b\right)^{\nu)\rho} \,,
\end{align}
where $p_a$ and $p_a^\prime$ denote the initial and final momenta of the particles and $\hat{n}$ is the unit vector on the celestial sphere. Here, the term ``particle'' is used in a broad sense \cite{Sen:2024qzb}, encompassing any asymptotic constituent, including macroscopic bodies and radiation. The coefficient $\tilde{h}_{\mu\nu}^{(0)}$ encodes the leading displacement memory~\cite{Christodoulou:1991cr,Thorne:1992sdb,He:2014laa,Strominger:2014pwa}, while $\tilde{h}_{\mu\nu}^{(\log)}$ encodes the subleading tail to the memory first derived systematically in~\cite{Sahoo:2018lxl,Saha:2019tub}. The sub-subleading coefficient $\tilde{h}_{\mu\nu}^{(\log)^2}$ was subsequently conjectured and verified in~\cite{Saha:2019tub,Sahoo:2020ryf}.

The logarithmic coefficient in eq.~\eqref{eq:sen_tom} has a useful decomposition before the massive-state rewriting.  The pairwise $S^{(1)}$ terms in $B_{\mu\nu}$ and $C_{\mu\nu}$ constitute the \emph{hard} contribution: they arise from the logarithmic deviations of the asymptotic hard-particle angular momenta.  The two terms in $B_{\mu\nu}$ proportional to $S^{(0)}$ constitute the \emph{drag} contribution.  With $P_+^\mu\equiv\sum_b p_b^{\prime\mu}$ and $P_-^\mu\equiv\sum_b p_b^\mu$, it is
\begin{equation} \label{eq:drag_coeff}
B^{\rm drag}_{\mu\nu}
=\frac{4G^2}{Rc^7}\left[
(P_+\!\cdot\!\hat n)S^{(0)}_{\mu\nu}(\{p'_a\},\hat n)
-(P_-\!\cdot\!\hat n)S^{(0)}_{\mu\nu}(\{p_a\},\hat n)
\right],
\qquad C^{\rm drag}_{\mu\nu}=0.
\end{equation}
It describes the backscattering of radiation through the long-range field of the hard system.  The full tail to the memory is the sum of the hard and drag pieces, $-(B_{\mu\nu}-C_{\mu\nu})$.  The split is tied to a choice of logarithmic frame: logarithmic translations redistribute terms between the drag phase and the hard-particle angular momenta, whereas their sum is invariant~\cite{Geiller:2024ryw,Boschetti:2025tru}.  The massless final-state terms cancel from the complete $B-C$ after the hard and drag pieces are combined and momentum conservation is used~\cite{Sahoo:2021ctw,Sen:2024qzb}.

\subsection{Memory effects and tails in the PN–MPM framework: consistency with soft theorems} \label{sec:memories_tails}

In this subsection we review how the universal soft-theorem coefficients $\tilde{h}_{\mu\nu}^{(0)}$, $\tilde{h}_{\mu\nu}^{(\log)}$, $\tilde{h}_{\mu\nu}^{(\log)^2}$ are realized within the MPM framework of Blanchet and Damour~\cite{Blanchet:1992br,Blanchet:2013haa}. The consistency between the PN--MPM hereditary waveform and the soft-theorem predictions provides the motivation and the benchmark for the NRGR amplitude computations in sections~\ref{sec:NRGR_memory} and~\ref{sec:higher}.

The GW-form is usually expressed in the transverse-traceless (TT) part of the asymptotic metric perturbation:
\be
h_{ij}^{\rm TT} = \Pi_{ijkl}\bar{h}^{kl} \,,
\ee
where
$\Pi_{ijkl}$ is the propagation direction-dependent projection operator, defined as
\be
\label{eq:proj_def}
\Pi_{ijkl}(\hat n) \equiv P_{ik}(\hat n)P_{jl}(\hat n) - \frac{1}{2} P_{ij}(\hat n)P_{kl}(\hat n) \,, \hspace{1cm} P_{ij}(\hat n) \equiv \delta_{ij} - \hat n_i \hat n_j \,,
\ee
where the unit vector $\hat{n}$ points from the source to the observer. When decomposed into a sum over \emph{radiative}
mass- and current-multipole moments:
\be \label{eq:radiative_tt}
h_{ij}^{\rm TT}(u,\hat n) =\ds \frac{4G}{R c^2} \Pi_{ijkl} \sum_{l=2}^\infty \frac{1}{c^l l!} \left[\mathcal{U}_{k l L-2} (u) \hat n_{L-2} + \frac{2l}{c(l+1)}\epsilon_{p q (k} \mathcal{V}_{l) p L-2} (u) \hat n_{q L-2}\right] + \mathcal{O}(R^{-2}) \,.\nonumber\\
\ee
The radiative mass- and
current-multipole moments $\mathcal{U}_{L} (u)$ and $\mathcal{V}_{L} (u)$ are symmetric trace-free (STF) tensors (with $L$ standing for $l$ STF spatial indices),  and are related to the \emph{source} multipole moments, which are defined in the near zone as integrals over the source. The explicit expression of the radiative mass quadrupole in terms of the intermediate canonical mass $\mathcal{M}_{ij}$ quadrupole moment is \cite{Blanchet:1997jj}:
\begin{align} \label{eq:radiative_moments}
\mathcal{U}_{ij}(u) = \mathcal{M}_{ij}^{(2)}(u) + \frac{2GM}{c^3}\int_{-\infty}^u \ln{\left(u-\tau\right)}\mathcal{M}_{ij}^{(4)} (\tau) d\tau -\frac{2}{7} \frac{G}{c^5} \int_{-\infty}^u \mathcal{M}_{a < i}^{(3)} (\tau) \mathcal{M}_{j > a}^{(3)} (\tau) d\tau \nonumber \\
 + \frac{2G^2 M^2}{c^6} \int_{-\infty}^u \ln^2 \left(u-\tau\right) \mathcal{M}_{ij}^{(5)} (\tau) d\tau + \mathcal{O}(c^{-7}) \,,
\end{align}
where we indicated the order of time derivatives with numbers in upper indices between round parentheses.

For the soft discussion, we have omitted the instantaneous 2.5PN terms, and the local-in-time contributions to the 1.5PN tail and 3PN tail-of-tail, which are analytic in frequency and do not affect the logarithmic hierarchy.\footnote{Consistently, we drop the constant quantity (with dimension) in the logarithm argument.
}

This expression therefore displays the three principal hereditary contributions involving only the lowest order radiative multipole, the mass-quadrupole: the 1.5PN tail $\sim (GM/c^3)\int\ln[(u-\tau)/2\tau_0]\,\mathcal{M}^{(4)}d\tau$~\cite{Blanchet:1987wq,Blanchet:1992br}, the 2.5PN nonlinear memory $\sim (G/c^5)\int\mathcal{M}^{(3)}\mathcal{M}^{(3)}d\tau$ \cite{Christodoulou:1991cr,Blanchet:1992br}, and the 3PN tail-of-tail $\sim (G^2M^2/c^6)\int\ln^2\mathcal{M}^{(5)}d\tau$ \cite{Blanchet:1997jj}.\footnote{Naming $L$ the size of the source and $v$ its internal velocity, the relations $\ddot {\cal M}_{ij}\sim Mv^2$ and $v^2\sim GM/L$ allow a straightforward power counting to identify the PN order of each term.} The tail-of-memory — the 4PN term $M\times {\cal M}_{ij}\times {\cal M}_{kl}$ — is not included in the compact formula (\ref{eq:radiative_moments}).

Note that the tail terms linear in the radiative multipole
(like the 1.5PN and the 3PN terms in (\ref{eq:radiative_moments})) scale respectively as $\omega \log(\omega)$, $(\omega\log\omega)^2$, and are thus sub-leading with respect to memory ones which are of the type $\omega^{n-1}\log^n(\omega)$.
The complete PN--MPM result contains instantaneous, tail-like, and genuine tail-of-memory contributions, while below we isolate its soft-logarithmic component~\cite{Trestini:2023ssa}.
The \emph{canonical} mass ($\mathcal M_L$) and momentum ($\mathcal S_L$) moments are related to the \emph{source}
ones $\mathcal{I}_L$ and $\mathcal{J}_L$ via
\be
\mathcal{M}_L = \mathcal{I}_L + G\delta\mathcal{I}_L + \mathcal{O}(G^2) \\
\mathcal{S}_L = \mathcal{J}_L + G\delta\mathcal{J}_L + \mathcal{O}(G^2) \,,
\ee

where, in the notation of e.g.~\cite{Blanchet:1998in}, the functions $\delta\mathcal{I}_L$ and $\delta\mathcal{J}_L$ depend on a set of six multipoles
$\{\mathcal{I}_L,\mathcal{J}_L,\mathcal{W}_L,\mathcal{X}_L,\mathcal{Y}_L,\mathcal{Z}_L\}$.
Among them, the mass and current moments $\mathcal{I}_L$ and $\mathcal{J}_L$ encode the physical multipolar content of the source, while $\mathcal{W}_L$, $\mathcal{X}_L$, $\mathcal{Y}_L$, and $\mathcal{Z}_L$ are gauge moments associated with the residual coordinate freedom in the MPM construction. They enter the nonlinear mapping from the source moments to the canonical moments; physical radiative observables are ultimately expressed in terms of the corresponding radiative moments.\footnote{E.g., at 2.5PN level ${\mathcal M}_{ij}={\mathcal I}_{ij}+\frac{2G}3\pa{I{\mathcal I}^{(3)}_{ij}- I^{(2)}{\mathcal I}^{(1)}_{ij}}$ where at leading order $ I^{(1)}=6{\mathcal W}$, with $I$ the moment of inertia of source, see (2.2) and (3.1) of \cite{Arun:2007sg} or \cite{Almeida:2024lbv}.}

The leading correction at 1.5PN order
is the mass-quadrupole tail, $M \times {\mathcal M}_{ij}$. At the 2.5PN
the leading non-linear memory
appears, due to the interaction between two quadrupole moments $\mathcal M_{ij} \times \mathcal M_{kl}$. At the 3PN order, the first term cubic in the source appears, called the tail-of-tail
and consisting of the interaction between two masses and the quadrupole
moment, i.e. $M^2 \times \mathcal M_{ij}$. The tail-of-memory appears at 4PN order, made of a cubic interaction
involving two time-varying quadrupole moments and the mass: $M \times \mathcal M_{ij}\times \mathcal M_{kl}$. In principle, it comes with a double time integration over the two
varying quadrupole moments $\sim \frac{G^2 M}{c^{8}} \int_{-\infty}^u d\tau_1 \, d\tau_2 \, \mathcal{M}_{a < i}^{(4)} (\tau_1) \mathcal{M}_{j > a}^{(4)} (\tau_1 -\tau_2) \ln{\left(\tau_2 \right)}$, unlike the memory term. 

Similarly, at 4.5PN order appears the first quartic interaction, $M^3 \times \mathcal M_{ij}$, naturally called the tail-of-tail-of-tail. The tail-of-tail-of-memory appears at 5.5PN order, made of a quartic interaction involving the two time-varying quadrupole moments and the two masses, i.e., $M^2 \times \mathcal M_{ij} \times \mathcal M_{kl}$. Dimensionally, it can be of the form $\sim \frac{G^3 M^2}{c^{11}} \int_{-\infty}^u d\tau_1 \, d\tau_2 d\tau_3 \, \, \mathcal{M}_{a < i}^{(5)} (\tau_1 -\tau_2) \mathcal{M}_{j > a}^{(5)} (\tau_1 -\tau_3) \ln{\left(\tau_2\right)} \ln{\left({\tau_3}\right)}$.

All tails, memory, and tail-of-memory are hereditary corrections to the waveform.

The soft counting, i.e the $\omega$ power counting,  is also straightforward. The memory starts as $1/\omega$. Each additional scattering of the emitted radiation off the static mass contributes a factor $GM\,\omega\,\log(\omega)$, where $\log(\omega)$ contains the infrared logarithm. It is useful to define an STF tensor
\begin{equation}
\tilde{\mathcal{Q}}_{ij}(\omega) \equiv
G\int\frac{d\omega'}{2\pi}\,(\omega-\omega')^3\omega'^3\,
\tilde{\mathcal{M}}_{a\langle i}(\omega-\omega')
\tilde{\mathcal{M}}_{j\rangle a}(\omega')\,,
\label{eq:memory_source_tensor}
\end{equation}
in terms of which, suppressing overall numerical coefficients, the soft hierarchy is
\begin{align}
\tilde{\mathcal{U}}_{ij}^{\mathrm{mem}}(\omega)&\sim \omega^{-1}\tilde{\mathcal{Q}}_{ij}(\omega)\,,
\\
\tilde{\mathcal{U}}_{ij,\mathrm{soft}}^{{\rm t}^n\rm{om}}(\omega)&\sim (GM)^n\omega^{n-1}[\log(\omega)]^n\tilde{\mathcal{Q}}_{ij}(\omega)\,, \nonumber
\end{align}
Below, we illustrate this explicitly for the memory and tail-of-memory terms using Fourier transforms.

\begin{itemize}
    \item \emph{Memory:} The non-linear memory at 2.5PN order is given by \cite{Blanchet:1997ji}
    \be \label{eq:memory}
    \mathcal{U}_{ij, 2.5 PN}^{\rm{mem}} (u) = -\frac{2G}{7 c^5} \int_0^\infty \, d\tau \, \mathcal{M}_{a<i}^{(3)}(u-\tau) \mathcal{M}_{j>a}^{(3)}(u-\tau) \,.
    \ee
    Using the inverse Fourier transform, we can rewrite the expression as
    \begin{align}
    \mathcal{U}_{ij, 2.5 PN}^{\rm{mem}} (u) = -\frac{2G}{7 c^5} \int_0^\infty \, d\tau \, \int \frac{d\omega_1}{2\pi} \frac{d\omega_2}{2\pi} \, e^{-i(\omega_1 + \omega_2)(u-\tau)} \, (-i\omega_1)^3 (-i\omega_2)^3 \tilde{\mathcal{M}}_{a<i}(\omega_1) \tilde{\mathcal{M}}_{j>a}(\omega_2) \,,
    \end{align}
and performing the $\tau$-integral, we get
    \be
    \mathcal{U}_{ij, 2.5 PN}^{\rm{mem}} (u) =i\frac{2G}{7c^5} \int \frac{d\omega_1}{2\pi} \frac{d\omega_2}{2\pi} \, e^{-i(\omega_1 + \omega_2)u} \, \frac{\omega_1^3 \omega_2^3}{(\omega_1 +\omega_2)}  \, \tilde{\mathcal{M}}_{a<i}(\omega_1) \tilde{\mathcal{M}}_{j>a}(\omega_2) \,.
    \ee
Writing the memory in frequency space:
    \be
    \nonumber
    \pa{\frac{2G}{7c^5}}^{-1}\tilde{\mathcal{U}}_{ij, 2.5 PN}^{\rm{mem}} (\omega) &=&\ds \int du \, e^{i\omega u} \,  \mathcal{U}_{ij, 2.5 PN}^{\rm{mem}} (u) \\ \nonumber
    &= &\ds \frac i{2\pi}\int d\omega_1 d\omega_2 \, \delta (\omega - \omega_1 -\omega_2) \, \frac{\omega_1^3 \omega_2^3}{(\omega_1 +\omega_2)}  \, \tilde{\mathcal{M}}_{a<i}(\omega_1) \tilde{\mathcal{M}}_{j>a}(\omega_2) \\
     &= &\ds \frac i{\omega} \int \frac{d\omega^\prime}{2\pi} \, (\omega^\prime)^3 (\omega - \omega^\prime)^3  \, \tilde{\mathcal{M}}_{a<i}(\omega^\prime) \tilde{\mathcal{M}}_{j>a}(\omega - \omega^\prime) \,.
    \ee
    
This gives $1/\omega\times$ the radiated energy (integral of the GW luminosity), and is consistent with the correct leading soft behaviour $\frac{1}{\omega}$.
\item \emph{Tail-of-memory:} The tail-of-memory appears at 4PN order \cite{Trestini:2023wwg}: 
    \begin{align}
    \mathcal{U}_{ij, 4 PN}^{\rm tom}(u)=\frac{8 G^2M}{7 c^8} \int_0^\infty \, d\rho \, \mathcal{M}_{a<i}^{(4)}(u-\rho) \int_0^\infty \, d\tau \,\mathcal{M}_{j>a}^{(4)}(u-\rho-\tau)  \ln{\left(\tau\right)}\,.
    \label{eq:tom_TD}
    \end{align}
Writing in the frequency domain:
    \begin{align}
    \nonumber
    \tilde{\mathcal{U}}_{ij, 4 PN}^{\rm tom} (\omega) &=\frac{8G^2M}{7c^8}  \int_0^\infty \, d\rho \int_0^\infty \, d\tau \ln{\left(\tau\right)} \int du \, e^{i\omega u} \\
    & \, \int \frac{d\omega_1}{2\pi} \frac{d\omega_2}{2\pi}\, e^{-i\omega_1(u-\rho)} \, e^{-i\omega_2(u-\rho-\tau)}\, (-i\omega_1)^4 (-i\omega_2)^4 \tilde{\mathcal{M}}_{a<i}(\omega_1) \tilde{\mathcal{M}}_{j>a}(\omega_2) \,,
    \end{align}
then performing the $\rho$, $\tau$ and $u$-integrals, we get
    \be
    \label{eq:tom}
    \ba{rcl}
    \ds\pa{\frac{8G^2M}{7c^8}}^{-1}\tilde{\mathcal{U}}_{ij, 4 PN}^{\rm tom} (\omega) &\simeq &\ds 
    \int \frac{d\omega_1}{2\pi} d\omega_2 \, \delta (\omega - \omega_1 -\omega_2) \, \frac{\omega_1^4 \omega_2^4}{\omega_2 (\omega_1 +\omega_2)}\\
    &&\ds\quad\times \ln\pa{\omega_2} \, \tilde{\mathcal{M}}_{a<i}(\omega_1) \tilde{\mathcal{M}}_{j>a}(\omega_2) \\
    &= & \ds\frac{1}\omega \int \frac{d\omega^\prime}{2\pi} \, (\omega^\prime)^4 (\omega - \omega^\prime)^3  \, \ln{(\omega - \omega^\prime)} \tilde{\mathcal{M}}_{a<i}(\omega^\prime) \tilde{\mathcal{M}}_{j>a}(\omega - \omega^\prime) \,,
    \ea
    \ee
    where in the approximate equality we have neglected non-logarithmic, $\omega$ independent terms in the result of the $\tau$ integration.
It is now convenient to define the symmetric memory kernel
\begin{equation}
\mathcal K_{ij}(\omega_1,\omega_2)
\equiv \omega_1^3\omega_2^3\,
\tilde{\mathcal M}_{a\langle i}(\omega_1)
\tilde{\mathcal M}_{j\rangle a}(\omega_2)\,,
\label{eq:memory_kernel}
\end{equation}
in terms of which we can rewrite eq.~(\ref{eq:tom}) as
\begin{equation} 
\tilde{\mathcal U}_{ij,4\,\mathrm{PN}}^{\mathrm{tom}}(\omega) =  \frac{8G^2M}{7c^8}\times\frac{1}{2\omega}\int\frac{d\omega^\prime}{2\pi}\,
\mathcal K_{ij}(\omega^\prime,\omega-\omega^\prime)
\left[\omega^\prime\ln|\omega-\omega^\prime|
+(\omega-\omega^\prime)\ln|\omega^\prime|\right].
\label{eq:tom_symmetrized}
\end{equation}
Now we split the integral into two regions: the hard region where $\vert\omega^\prime\vert \gg \omega$, and the soft region where $\vert \omega^\prime\vert \sim \omega$. While the hard-region contribution is analytic in $\omega$, the soft region has the unique nonanalytic contribution $\omega\ln|\omega|$. The prefactor $1/\omega$ in
\eqref{eq:tom_symmetrized}, together with $d\omega'=\omega\,dx$, converts
this kernel contribution into the $\ln|\omega|$ tail of the radiative
quadrupole:
\begin{equation}
\left.\tilde{\mathcal U}_{ij,4\,\mathrm{PN}}^{\mathrm{tom}}(\omega)
\right|_{\log|\omega|}
\propto \frac12\ln|\omega|\int_0^1\frac{dx}{2\pi}\,
\mathcal K_{ij}\bigl(\omega x,\omega(1-x)\bigr)\,,
\label{eq:tom_soft_result}
\end{equation}
which shows the desired non-analytic dependence on $\omega$; see
Appendix~\ref{sec:app_memory_angular} for detailed derivation. 

Similarly, for tail-of-tail-of-memory at 5.5 PN, the analysis involves three time integrals and schematically, in the soft region we get
    \be
    \tilde{\mathcal{U}}_{ij, 5.5 PN}^{{\rm t}^2\rm{om}} (\omega) \propto \frac{(\ln{\omega})^2}{\omega} \int d\omega^\prime \, (\omega^\prime)^3 (\omega - \omega^\prime)^3  \, \omega^2 \tilde{\mathcal{M}}_{a<i}(\omega^\prime) \tilde{\mathcal{M}}_{j>a}(\omega - \omega^\prime) \,,
    \ee
consistent with the sub-subleading soft behavior $\omega (\ln{\omega})^2$. This pattern continues to higher orders where the general form of the \emph{universal} soft behaviour is $\omega^m (\ln{\omega})^{m+1}$.  

In addition to the tail-of-tail and the tail-of-memory, there is also a cubic interaction involving the constant angular momentum, i.e. $M \times S_i \times {\mathcal M}_{jk}$, called the \emph{spin-quadrupole tail} which appears at 4PN order \cite{Trestini:2023wwg}. Its associated spin-tail-of-memory is expected to appear at 6.5 PN order, made of a quartic interaction involving two time-varying quadrupole moments, the mass and the angular momentum, $M \times S_i \times \mathcal M_{jk} \times \mathcal M_{pq}$. By ``dimensional'' analysis it is expected to be of the form $\sim \frac{G^3 M}{c^{13}} S_b \epsilon_{b c a } \int_{-\infty}^u d\tau_1 \int^{\tau_1}_{-\infty} d\tau_2 \, \mathcal{M}_{c < i}^{(5)} (\tau_1) \mathcal{M}_{j > a}^{(5)} (\tau_1 -\tau_2) \ln{\left(\tau_2 \right)} $,
and represents a sub-leading, non-universal correction to the tail of memory.

\end{itemize}

\subsection{Non-linear gravitational memory: PM $\rightarrow$ PN-MPM approximation}
\label{ssec:nonlinmem}

In this subsection we derive the non-linear (Christodoulou) memory formula by starting from the leading soft theorem at the PM level and reducing it to the PN--MPM quadrupole approximation~\cite{Christodoulou:1991cr,Wiseman:1991ss,Thorne:1992sdb,Blanchet:1992br}. 
The resulting expression matches the 2.5PN nonlinear memory term in eq.~\eqref{eq:radiative_moments}, providing a direct bridge between the soft-theorem and the PN framework.

The small frequency limit of the radiative field at null infinity is
\be
\label{eq:hTTsoft}
\ba{rcl}
\tilde h^{\rm TT}_{ij} (\omega, \hat{n}) &\overset{\omega\to 0}{=}
& \ds\frac{4G}{R c^4}  \Pi_{ijkl}(\hat n)\sum_a \frac{p_a^k p_a^l}{p_a \cdot q} \\
&=&\ds\frac{4G}{R c^4} \frac{1}{\omega} \Pi_{ijkl}(\hat n) \sum_a \frac{p_a^k p_a^l}{(E_a  - \mathbf{p}_a \cdot \hat{n})}\,,
\ea
\ee
where we have used $p_a^{\mu} = (E_a, \mathbf{p}_a) \,, p_a \cdot q = \omega (E_a  - \mathbf{p}_a \cdot \hat{n})$. 

The memory effect gets its name from the non-vanishing difference between the field at late and early retarded times, $\Delta h_{ij} (r,\hat{n}) \equiv \lim_{u \to +\infty} h^{\rm TT}_{ij} (u,r,\hat n) - \lim_{u \to -\infty} h^{\rm TT}_{ij} (u,r,\hat n)$; the $\frac{1}{\omega}$ pole is (anti-)Fourier transformed into a step function in retarded time $u$. We obtain
\be
h_{ij}^{\rm TT} (u,\hat{n}) \overset{\omega\to 0}= \frac{4G}{R c^4} \Pi_{ijkl}(\hat n) \sum_a \left[\frac{p_a^k p_a^l}{(E_a  - \mathbf{p}_a \cdot \hat{n})} \right]_{\text{in}}^{\text{out}} \Theta (u)\,.
\ee

This is the leading soft theorem memory formula, holding for "hard" external states, both massive and massless particles, including hard
gravitons. For linear memory, the radiative field is sourced by the matter particles, but for non-linear memory the sources are the hard gravitons as they themselves emit further gravitons. Therefore, replacing the discrete sum over hard gravitons of \eqref{eq:hTTsoft} by a classical wave packet, we get
\be
\label{eq:hmemE}
h_{ij}^{\rm TT} (u,\hat{n}) \overset{\omega\to 0}= \frac{4G}{R c^4 }  \Pi_{ijkl}(\hat{n})\int_{-\infty}^u du^\prime \int d\Omega^\prime \,   \frac{\hat{n}_k^\prime \hat{n}_l^\prime}{(1  - \hat{n}^\prime \cdot \hat{n})} \frac{dE}{du^\prime d\Omega^\prime} \,,
\ee
where we have parametrized the momenta of hard gravitons as $p_a = \omega' (1,\hat{n}^\prime)$. 

The differential luminosity per solid angle carried by GWs is
\begin{equation}
\frac{dE}{du\,d\Omega}
=\frac{c^3R^2}{32\pi G}\,
\dot h_{ij}^{\rm TT}(u,\hat n)\dot h_{ij}^{\rm TT}(u,\hat n)\,.
\label{eq:energy_flux_from_waveform}
\end{equation}
At leading quadrupole order, the waveform entering this formula is
\begin{equation}
\left. h_{ij}^{\rm TT}(u,\hat n)\right|_{\rm Quad}
=\frac{2G}{Rc^4}\,
\Pi_{ijkl}(\hat n)\mathcal M_{kl}^{(2)}(u)\,.
\label{eq:quadrupole_waveform_energy_flux}
\end{equation}
Inserting this waveform into~\eqref{eq:energy_flux_from_waveform} gives the quadrupolar contribution to the luminosity
\begin{equation}
\left.\frac{dE}{du\,d\Omega}\right|_{\rm Quad}
=\frac{G}{8\pi c^5}\,
\Pi_{ijkl}(\hat n)\,
\mathcal M_{ij}^{(3)}(u)\mathcal M_{kl}^{(3)}(u)\,.
\label{eq:quadrupole_energy_flux_projector}
\end{equation}

The differential luminosity admits an STF expansion, which for \eqref{eq:quadrupole_energy_flux_projector} takes the form
\be
\left. \frac{dE}{du \, d\Omega}\right|_{\rm Quad} = \sum_{\ell=0,2,4} \mathcal{F}_{L}(u) \hat n_L\,,
\ee
defining the newly introduced $\mathcal F_L$, which are STF projections of the differential luminosity. To STF-expand the contraction in \eqref{eq:quadrupole_energy_flux_projector}, we write
\begin{displaymath}
\begin{aligned}
\Pi_{ijkl}(\hat n)\mathcal M_{ij}^{(3)}\mathcal M_{kl}^{(3)}
={}&\mathcal M_{ij}^{(3)}\mathcal M_{ij}^{(3)}
-2\mathcal M_{ai}^{(3)}\mathcal M_{aj}^{(3)}\hat n_i\hat n_j
+\frac12\bigl(\mathcal M_{ij}^{(3)}\hat n_i\hat n_j\bigr)^2\,,
\end{aligned}
\end{displaymath}
which can be STF-decomposed via\footnote{This follows from the following identities: 
\begin{displaymath}
\begin{array}{rcl}
\hat n_i\hat n_j&=&\hat n_{\langle ij\rangle}+\frac13\delta_{ij},\label{eq:unit_vector_rank2_stf}\\
\hat n_i\hat n_j\hat n_k\hat n_l
&=&\hat n_{\langle ijkl\rangle}
+\frac17\Big(\delta_{ij}\hat n_{\langle kl\rangle}
+\delta_{ik}\hat n_{\langle jl\rangle}
+\delta_{il}\hat n_{\langle jk\rangle}
+\delta_{jk}\hat n_{\langle il\rangle}
+\delta_{jl}\hat n_{\langle ik\rangle}
+\delta_{kl}\hat n_{\langle ij\rangle}\Big)\\
&&\quad+\frac1{15}\Big(\delta_{ij}\delta_{kl}
+\delta_{ik}\delta_{jl}+\delta_{il}\delta_{jk}\Big).
\end{array}
\end{displaymath}}
\be
\begin{aligned}
\mathcal M_{ai}^{(3)}\mathcal M_{aj}^{(3)}\hat n_i\hat n_j
={}&\mathcal M_{a<i}^{(3)}\mathcal M_{j>a}^{(3)}
\hat n_{<ij>} +\frac13\mathcal M_{ab}^{(3)}\mathcal M_{ab}^{(3)},\\
\bigl(\mathcal M_{ij}^{(3)}\hat n_i\hat n_j\bigr)^2
={}&\mathcal M_{<ij}^{(3)}\mathcal M_{kl>}^{(3)}
\hat n_{<ijkl>} +\frac47\mathcal M_{a<i}^{(3)}
\mathcal M_{j>a}^{(3)}\hat n_{<ij>}
+\frac{2}{15}\mathcal M_{ab}^{(3)}\mathcal M_{ab}^{(3)} .
\end{aligned}
\label{eq:FL}
\ee
Thus the flux contains $\ell=0,2,4$ components.  Its $\ell=2$ projection is
\be
\left.\frac{dE}{du\,d\Omega}\right|_{{\rm Quad},\ell=2}
=\frac{G}{8\pi c^5}\,\mathcal F_{ij}(u)\hat n_{<ij>},
\qquad
\mathcal F_{ij}(u)= -\frac{12}{7}\,
\mathcal M_{a<i}^{(3)}(u)\mathcal M_{j>a}^{(3)}(u).
\label{eq:quadrupole_flux_l2}
\ee
Here the coefficient $-12/7=-2+\frac12(4/7)$ follows directly from the
two preceding STF decompositions.  The $\ell=0$ term gives
$\frac{G}{5c^5}\mathcal M_{ab}^{(3)}\mathcal M_{ab}^{(3)}$ after angular
integration, as required by the quadrupole luminosity formula and gives a vanishing contribution to the memory integral \eqref{eq:hmemE}; the
$\ell=4$ term contributes to the radiative mass hexadecapole, not to the
radiative mass quadrupole. The corresponding angular integrals are given
in Appendix~\ref{sec:app_memory_angular}.

Inserting \eqref{eq:quadrupole_flux_l2} into the PM memory formula \eqref{eq:hmemE} gives the quadrupolar contribution to the memory
\be
\left.h_{ij}^{\rm mem}(u,\hat n)\right|_{\rm{Quad},\ell=2}
=\frac{G^2}{2\pi R c^9}
\Pi_{ijkl}(\hat n)
\int_{-\infty}^{u}du'\,
{\mathcal F}_{mn}(u')
\int d\Omega'\,
\frac{\hat n_k'\hat n_l'\hat n_{<mn>}'}{1-\hat n\cdot\hat n'}.
\label{eq:memory_flux_angular_integral}
\ee

Combining the result
\be
\Pi_{ijkl}(\hat n)
\int d\Omega'\,
\frac{n_k'n_l'}{1-\hat n\cdot\hat n'}\,
n_{<mn>}'
=\frac{2\pi}{3}\Pi_{ijmn}(\hat n)\,,
\label{eq:memory_kernel_l2}
\ee
which is derived in
Appendix~\ref{sec:app_memory_angular}, with  \eqref{eq:quadrupole_flux_l2}, we obtain

\be
\begin{aligned}
\left.h_{ij}^{\rm mem}(u,\hat n)\right|_{{\rm Quad},\ell=2}
&=\frac{G^2}{3R c^9}\Pi_{ijkl}(\hat n)\int_{-\infty}^{u}du'\,
\mathcal F_{kl}(u')\\
&=-\frac{4G^2}{7R c^9}\Pi_{ijkl}(\hat n)
\int_{-\infty}^{u}du'\,
\mathcal M_{a<k}^{(3)}(u')\mathcal M_{l>a}^{(3)}(u')\\
&=\frac{2G}{R c^4}\Pi_{ijkl}(\hat n)\mathcal U_{kl}(u).
\end{aligned}
\ee
Comparing with eq. \eqref{eq:radiative_tt}, we obtain
\be \label{eq:exact_memory}
\mathcal{U}^{{\rm mem}}_{ij,2.5PN}(u) = -\frac{2G}{7 c^5}\int_{-\infty}^{u}d\tau\,
\mathcal M_{a<i}^{(3)}(\tau)\mathcal M_{j>a}^{(3)}(\tau)\,,
\ee
which is equivalent to \eqref{eq:memory}.
This can be interpreted as a modification of the mass-type radiative quadrupole~\cite{Blanchet:2013haa}:
\be
\left.{\mathcal U}_{ij}^{\rm mem} (u)\right|_{\rm Quad} = -\frac{2G}{7c^5} \int_{-\infty}^{u} d\tau \, \mathcal{U}_{a<i}^{(1)} (\tau) \, \mathcal{U}_{j>a}^{(1)} (\tau) \,,
\ee
which, using
the expression of the radiative mass quadrupole in eq.~\eqref{eq:radiative_moments}, gives
\be \label{eq:chain_of_memories}
\ba{rcl}
\left.{\mathcal U}_{ij}^{\rm mem} (u)\right|_{\rm Quad} &=& \underbrace{-\frac{2G}{7c^5} \int_{-\infty}^u d\tau \, \mathcal{M}_{a<i}^{(3)} (\tau) \, \mathcal{M}_{j>a}^{(3)} (\tau) }_{\text{2.5PN memory}} \\
&-& \underbrace{\frac{8G^2 M}{7 c^8} \int_{-\infty}^u d\tau_1 \mathcal{M}_{a<i}^{(3)} (\tau_1) \, \int_0^{\infty} d\tau_2 \, \mathcal{M}_{j>a}^{(5)} (\tau_1 -\tau_2) \ln{\left(\tau_2\right)} }_{\text{4PN tail-of-memory}} \\
&-& \underbrace{\frac{8G^3 M^2}{7 c^{11}} \int_{-\infty}^u d\tau_1 \int_0^\infty d\tau_2\, \mathcal{M}_{a<i}^{(5)} (\tau_1 -\tau_2)\ln{\left(\tau_2\right)} \int_0^\infty \,d\tau_3 \, \mathcal{M}_{j>a}^{(5)} (\tau_1 -\tau_3) \, \ln{\left(\tau_3\right)} }_{\text{5.5PN tail-of-tail-of-memory}} \\\ &+& \, \mathcal{O}\left(\frac{1}{c^{14}}\right) \,.
\ea
\ee
The three labelled terms form the beginning of the hereditary hierarchy in the soft limit: the 2.5PN nonlinear memory, the 4PN tail-of-memory (one mass insertion), and the 5.5PN tail-of-tail-of-memory (two mass insertions). Each successive term involves an additional time convolution and one additional factor of $2GM/c^3$, corresponding to one additional scattering of the radiation off the static Schwarzschild curvature. In frequency space, these successive integrations translate directly into the soft hierarchy $1/\omega\to\log\omega\to\omega(\log\omega)^2$, as we verified explicitly.\footnote{We verified the exact numerical coefficient up to the tail-of-memory term.}

The relation of the 4PN term to the drag part of the logarithmic soft theorem can be stated directly.  In the static-monopole limit of the drag contribution \eqref{eq:drag_coeff}, the non-analytic correction factorizes into the radiative leading-soft field times the long-distance transfer function
\begin{equation}
\widetilde h^{(\log),\mathrm{drag}}_{ij}(\omega,\hat n)|_{\log}
\simeq
\mathcal T_M^{\log}(\omega)\,
\widetilde h^{\mathrm{mem,rad}}_{ij}(\omega,\hat n),
\qquad
\mathcal T_M^{\log}(\omega)
\equiv\frac{2iGM\omega}{c^3}\log|\omega|.
\label{eq:soft_tail_transfer}
\end{equation}
Here $\widetilde h^{\mathrm{mem,rad}}\propto 1/(\omega+i0^+)$ is the radiative, null-memory part of the leading soft field.  The $1.5$PN MPM tail has the same non-analytic transfer function; its finite constant and phase overall shift depend on the choice of radiative time and are not fixed by~\eqref{eq:soft_tail_transfer}.  Equation~\eqref{eq:soft_tail_transfer} therefore maps the leading $1/\omega$ memory behavior into the drag contribution proportional to $\log\omega$, and it is realized multipolarly in eqs.~\eqref{eq:exact_memory}(see ~\eqref{eq:radiative_moments} for the tail and tail$^2$  contributions linear in the radiative multipole): the tail transfer function can dress either radiative quadrupole in the quadratic memory flux.  
Thus the soft drag factor and the PN--MPM tail-of-memory describe the same long-distance radiative mechanism.

Although the linear memory formula is structurally leading order (0PN), its physical relevance depends entirely on the system’s boundary conditions. For \emph{unbound} systems, such as hyperbolic scattering, it appears at leading order in the waveform, scaling as the space-space component of the energy-momentum tensor $T_{ij}\sim (v^2/c^{2})$. For \emph{bound} binaries, however, the linear memory vanishes identically unless the merger ejects mass or the final remnant receives a recoil (kick). For nonspinning binaries, the recoil velocity scales as $v_{\rm recoil}\sim (v/c)^{7}$, corresponding to a 3.5PN effect, as can be verified by integrating the linear momentum flux formula; see e.g.~(35) of \cite{Blanchet:2013haa}. Since the linear memory is quadratic in the recoil velocity, the resulting kick memory is suppressed to much higher PN order. Consequently, the nonlinear (Christodoulou) memory dominates for binary mergers; sourced by the GW energy flux, it scales as $(v/c)^9$, i.e., 2.5PN relative to the leading-order waveform $(v/c)^4$. Recoil memory is subleading because it arises from the radiative backreaction on the source, rather than from a direct change in external asymptotic states as in hyperbolic scattering. Unlike the hyperbolic case, any change in velocity (recoil) in a bound system is strictly accompanied by a corresponding change in the energy flux. 

In the PN approximation scheme, tails-of-memory and their higher-order tails are hereditary logarithmic corrections to the leading-order memory, arising from the scattering of gravitational radiation off the long-range $1/r$ Coulombic field of the source. In soft-theorem language, these effects correspond to infrared
logarithms generated by long-range gravitational interactions in the low-frequency expansion of radiative fields.

\subsection{Spin memory and the subleading soft theorem}

Spin memory is the magnetic-parity counterpart of displacement memory.  Whereas
displacement memory is a change in the strain sourced by energy flux, spin
memory is an enduring change in the magnetic-parity part of the \emph{retarded-time
integral} of the strain. More precisely, it is the magnetic-parity part
of $ \int du \, h_{AB}^{\rm TT}$, where $A,B$ are indices parametrizing directions tangent to the unit two-sphere. Operationally, it can be measured as the relative
arrival-time delay of counter-orbiting light rays around a large closed
contour, and is therefore distinct from the permanent displacement measured
by the ordinary memory~\cite{Pasterski:2015tva,Nichols:2017rqr}.

The angular-momentum-flux origin of this observable is the classical counterpart
of the ordinary subleading soft graviton theorem. In the convention used here,
the latter takes the form
\begin{equation}
S^{(1)}(\varepsilon,q)=-\frac{i\kappa}{2}\sum_a\eta_a\,
\frac{\varepsilon_{\mu\nu}q_\rho J_a^{\rho\mu}p_a^\nu}
{p_a\cdot q}\,,
\label{eq:subleading_soft_spin}
\end{equation}
where $J_a^{\mu\nu}$ is the total angular momentum of hard particle $a$ and
 $\eta_a=+1$ ($-1$) for outgoing (incoming) particles.  The time-domain
representation of this soft factor gives the angular-momentum-flux representation
of spin memory~\cite{Pasterski:2015tva}.
The radiative field associated with the sub-leading soft factor is
\begin{equation}
\widetilde h_{ij}^{(1)}(\omega,\hat n)
=-\frac{2G}{Rc^4}\,\Pi_{ijkl}(\hat n)
\sum_a\eta_a\,
\frac{q_\rho J_a^{\rho(k}p_a^{l)}}{p_a\!\cdot q}\,.
\label{eq:subleading_soft_waveform}
\end{equation}
For massless hard gravitons this becomes, after replacing the discrete sum by
the angular-momentum flux,
and specializing to space-space part of $J^{\rho k}$
\begin{equation}
\int_{-\infty}^{+\infty}\!du\,h_{ij}^{\mathrm{TT,spin}}(u,\hat n)
=-\frac{2G}{Rc^4}\,\Pi_{ijkl}(\hat n)
\int d\Omega'\,
\frac{\hat n_m \hat n'^{l}}
{1-\hat n\!\cdot\hat n'}
\int_{-\infty}^{+\infty}\!du\,
\frac{dJ^{mk}}{du\,d\Omega'}\,.
\label{eq:spin_memory_soft_flux_waveform}
\end{equation}
This finite, $\omega^0$ term is the retarded-time integral of the strain.
However, being analytic in $\omega$
it is expected that analogous terms may arise from local dynamics. While ``non-analytic terms cannot come from the integral of a finite quantity over a finite region'', analytic ones can receive
contributions from local interactions over a finite region around the scattering center \cite{Sen:2024bax}; thus, we do not expect the spin-memory kernel \eqref{eq:subleading_soft_waveform} to match the PN-MPM result as for the soft non-linear energy memory in section~\ref{ssec:nonlinmem}.

Following the treatment of the non-linear memory,
the general formula for the differential angular momentum emission, at leading quadrupole order, is, see e.g. section 3.3.3 of \cite{Maggiore:2007ulw}:
\begin{align}
\left.\frac{dJ^{mk}}{du\,d\Omega}\right|_{\rm Quad}
=&\frac{G}{8\pi c^5}\,
{\mathcal M}_{ab}^{(3)}{\mathcal M}_{cd}^{(2)}
\bigg\{\hat n^{m}\pa{\hat n_{c}\Pi_{abkd}
+\hat n_{d}\Pi_{abkc}}
 +2
\Pi_{pmab}\Pi_{pkcd}\bigg\}-m\leftrightarrow k\,,
\label{eq:ang_momentum_flux_density}
\end{align}
where the argument of the projectors is understood to be $\hat n$. 

The $\ell=1$ projection of the differential angular momentum flux is
\be
\left.\frac{dJ^{mk}}{du\,d\Omega}\right|_{{\rm Quad},\ell=1}
=\frac{G}{10\pi c^5} \left(\mathcal M_{ma}^{(2)}\mathcal M_{ak}^{(3)}
-\mathcal M_{ma}^{(3)}\mathcal M_{ak}^{(2)} \right)\,,
\label{eq:ang_momentum_flux_L0}
\ee
and its sphere integral gives
\be
\left.\frac{dS_i}{du}\right|_{{\rm Quad},\ell=1}
=\frac{2G}{5c^5}\epsilon_{imk}
\mathcal M_{ma}^{(2)}\mathcal M_{ka}^{(3)},
\label{eq:quadrupolar_ang_momentum_luminosity}
\ee
which is the standard leading order (quadrupole) angular-momentum luminosity \cite{Maggiore:2007ulw}, when expressed in terms of the angular momentum pseudo-vector $S_i\equiv \frac 12\epsilon_{ijk}J^{jk}$, but which does not contribute to the spin-memory integral.
The first non-vanishing contribution to the angular integral in \eqref{eq:spin_memory_soft_flux_waveform} is the $\ell=3$ projection, which can be written
in terms of
\begin{align}
\mathcal J^{mk}_{ab}\equiv{}&
\mathcal M_{kp}^{(2)}
\mathcal M_{p\langle a}^{(3)}\delta_{b\rangle m}
-\mathcal M_{m\langle a}^{(2)}\mathcal M_{b\rangle k}^{(3)}
-m\leftrightarrow k\,.
\label{eq:spin_memory_Jmkab}
\end{align}
as
\begin{equation}
\left.\frac{dJ^{mk}}{du\,d\Omega}\right|_{{\rm Quad},\ell=3}
=\frac{3G}{14\pi c^5}\mathcal J^{mk}_{ab}(u)
\hat n_{\langle ab\rangle}.
\label{eq:ang_momentum_flux_L2}
\end{equation}
To perform the angular integration in \eqref{eq:spin_memory_soft_flux_waveform} we can use; see the derivation in 
Appendix~\ref{sec:app_spin_memory_angular}, 
\begin{equation}
\Pi_{ijkl}(\hat n)\int d\Omega'\,
\frac{\hat n'^l \hat n'_{\langle ab\rangle}}
{1-\hat n\!\cdot\hat n'}
=\frac{4\pi}{3}\Pi_{ijk\langle a}(\hat n)\hat n_{b\rangle},
\label{eq:spin_memory_kernel_L2}
\end{equation}
which yields the contribution to the spin-memory from the $\ell=3$, ${\mathcal M}_{ij}\times {\mathcal M}_{ij}$
\begin{align}
\int_{-\infty}^{+\infty}\!du\,
h_{ij}^{\mathrm{TT,spin,Q}}(u,\hat n)\supset {}&-\frac{3G^2}{7\pi Rc^9}\Pi_{ijka}(\hat n)\hat n^m 
\int d\Omega'\frac{\hat n'^l\hat n'_{\langle ab\rangle}}{1-\hat n\cdot \hat n'}\int_{-\infty}^{+\infty} du {\mathcal J}^{mk}_{ab}(u)\nonumber\\
={}&-\frac{2G^2}{7Rc^9}\Pi_{ijka}(\hat n)
\int_{-\infty}^{+\infty}\!du\,\bigg[
\mathcal M_{kp}^{(2)}\mathcal M_{pa}^{(3)} \nonumber\\
&-\left(\mathcal M_{pq}^{(2)}\hat n_p\hat n_q\right)\mathcal M_{ka}^{(3)}
+\left(\mathcal M_{pq}^{(3)}\hat n_p\hat n_q\right)\mathcal M_{ka}^{(2)}
\bigg].
\label{eq:spin_memory_L2_final}
\end{align}

The remaining $l=5$ projection can be written in terms of
\be
{\mathcal J}^{mk}_{abcd}\equiv
\delta_{m\langle a}\mathcal M_{bc}^{(3)}\mathcal M_{ d\rangle k}^{(2)}
-m\leftrightarrow k
\label{eq:spin_memory_Qmkabcd}
\ee
as
\begin{equation}
\left.\frac{dJ^{mk}}{du\,d\Omega'}\right|_{{\rm Quad},\ell=5}
=-\frac{G}{8\pi c^5}{\mathcal J}^{mk}_{abcd}
\hat n'_{\langle abcd\rangle}.
\label{eq:ang_momentum_flux_L4}
\end{equation}
To perform the angular integration, we can use, see the derivation in Appendix~\ref{sec:app_spin_memory_angular},
\be
\Pi_{ijkl}(\hat n)\int d\Omega'\,
\frac{\hat n'^l \hat n'_{\langle abcd\rangle}}{1-\hat n\cdot\hat n'}
=\frac{4\pi}{5}\Pi_{ijk\langle a}(\hat n)\hat n_b\hat n_c\hat n_{d\rangle}\,,
\label{eq:spin_memory_kernel_L4}
\ee
which contributes to the spin-memory kernel as
\be
\begin{aligned}
\int_{-\infty}^{+\infty}du\,
h_{ij}^{\mathrm{TT,spin,Q}}\supset\frac{G^2}{Rc^9}\Pi_{ijka}(\hat n)
\int_{-\infty}^{+\infty}du\,\bigg[
&\frac{3}{35}\mathcal M_{kp}^{(2)}\mathcal M_{pa}^{(3)}
+\frac{9}{140}\left(\mathcal M_{pq}^{(2)}\hat n_p\hat n_q\right)\mathcal M_{ka}^{(3)}
\\
&+\frac{3}{35}\left(\mathcal M_{pq}^{(3)}\hat n_p\hat n_q\right)\mathcal M_{ka}^{(2)}
\bigg].
\end{aligned}
\label{eq:spin_memory_L4_final}
\ee
Combining the non-vanishing $\ell=3$ and $\ell=5$ sectors gives the complete
quadrupole--quadrupole soft spin-memory result,
\be
\begin{aligned}
\int_{-\infty}^{+\infty}du\,h_{ij}^{\mathrm{TT,spin,Q}}(u,\hat n)
=-\frac{G^2}{5Rc^9}\Pi_{ijka}(\hat n)
\int_{-\infty}^{+\infty}du\,\bigg[
&{\mathcal M}_{kp}^{(2)}{\mathcal M}_{pa}^{(3)}
-\frac74\left(\mathcal M_{pq}^{(2)}\hat n_p\hat n_q\right)\mathcal M_{ka}^{(3)}
\\
&+\left(\mathcal M_{pq}^{(3)}\hat n_p \hat n_q\right)\mathcal M_{ka}^{(2)}
\bigg]\,,
\end{aligned}
\label{eq:spin_memory_quadrupole_complete}
\ee
which contains both $\ell=2$ (first term inside square brackets on the r.h.s.) and $\ell=3$ terms (remaining ones inside the square bracket).
As commented earlier, and differently from the non-analytic non-linear memory, the soft spin ``memory'' contribution \eqref{eq:spin_memory_quadrupole_complete} is \emph{not} expected to match analog analytic terms quadratic in the mass-quadrupole in the PN-MPM formulae; see e.g., eqs. (5.10),  of \cite{Blanchet:1997ji} and (5.9) of \cite{Blanchet:2004bb} (and the discussion in section E of \cite{Almeida:2024lbv}) for mass quadrupole and (5.12) of  \cite{Blanchet:1997ji} for the current octupole:
\begin{align}
\left.h_{ij}^{TT,{\rm Q}}(u,\hat n)\right|_{\ell =2,3}=\Pi_{ijkl}(\hat n)\frac{G^2}{Rc^9}&\left[
\frac 97{\mathcal M}^{(5)}_{ka}{\mathcal M}_{aj}+
\frac{11}7{\mathcal M}^{(4)}_{ka}{\mathcal M}^{(1)}_{aj}+
\frac 67{\mathcal M}^{(3)}_{ka}{\mathcal M}^{(2)}_{aj}
\right.\nonumber\\
&\left.+\hat n_{\langle qa\rangle}
\epsilon_{pqk}\epsilon_{bc\langle l}\pa{
\frac1{10}\mathcal M_{p\underline{b}}^{(5)}
\mathcal M_{a\rangle c}-\frac 12{\mathcal M}_{p\underline{b}}^{(4)}
{\mathcal M}_{a\rangle c}^{(1)}}\right]\,.
\label{eq:httl23}
\end{align}

On the other hand, the PN result in the second line of \eqref{eq:httl23}, which can be rewritten as
\begin{align}
\left .h_{ij}^{TT,\rm{Q}}(u,\hat n)\right|_{\ell=3}=& \frac{G^2}{Rc^9}\Pi_{ijkl}(\hat n)\Big\{\frac 3{10}\paq{\pa{{\mathcal  M}^{(3)}_{ab}\hat n_a\hat n_b}{\mathcal M}^{(2)}_{kl}-\pa{{\mathcal  M}^{(2)}_{ab}\hat n_a\hat n_b}{\mathcal M}^{(3)}_{kl}}\nonumber\\
&+\frac d{dt}\paq{\frac 1{10}{\mathcal M}^{(4)}_{ka}{\mathcal M}^{(1)}_{la}-
\frac 35{\mathcal M}^{(3)}_{ka}{\mathcal M}^{(2)}_{la}}\Big\}\,,
\end{align}

can be related to the spin-memory by adding the $\ell=3$ contribution from \emph{superrotations} to the soft formula \eqref{eq:spin_memory_soft_flux_waveform}, as done in \cite{Nichols:2017rqr}.

We emphasize that this spin memory is distinct from the one discussed at the end of section \ref{sec:memories_tails}, the \emph{spin-dependent tail memory} \cite{Ghosh:2021bam}: the former belongs to the magnetic
parity/subleading-soft sector, while the latter is a subleading, non-universal correction to the tail of memory.

\section{Memory effect and tail-of-memory from amplitudes in NRGR} \label{sec:NRGR_memory}
 
We now turn to the explicit computation of the memory and tail-of-memory amplitudes within the NRGR effective field theory framework~\cite{Goldberger:2004jt,Almeida:2021jyt}, described in detail in Appendix~\ref{sec:app_nrgr}. Working in $d = 3-\epsilon$ spatial dimensions with retarded propagators and renormalization scale $\tilde{\mu}^2 = \pi\mu^2 e^{-\gamma}$, we evaluate the Feynman diagrams corresponding to each hereditary process and isolate their IR and UV divergence structure. Adopting the Kaluza-Klein decomposition of the metric polarization $g_{\mu\nu}\to\pag{\phi,A_i,\sigma_{ij}}$ \cite{Kol:2007bc}, the key organizing principle is that each static mass insertion $M$ (coupled to $\phi$) introduces one additional static propagator via the $\sigma\sigma\phi$ vertex, thereby generating one further power of the master integral $I_1(\omega)$ defined in \eqref{eq:master_integral_tail},
and consequently one more power of the hereditary logarithm $\log\omega$ in the soft limit. We first treat the memory at 2.5PN and the tail-of-memory at 4PN separately, then generalize to arbitrary $n$ tail insertions.
 
\subsection{The Memory Effect at 2.5PN}
\label{ssec:mem}
 
The nonlinear (Christodoulou) memory arises from the stress-energy tensor of the emitted gravitational radiation sourcing a further GW~\cite{Christodoulou:1991cr,Thorne:1992sdb}. In the NRGR framework, this process has been studied in \cite{Almeida:2024lbv} and it is represented by a one-loop diagram in which two mass quadrupole sources radiate into the same external graviton. Its TT radiative component is finite (with no IR or UV pole) and produces only the Weinberg $1/\omega$ soft pole, which is the PN realisation of the leading soft coefficient $\tilde{h}_{\mu\nu}^{(0)}$ in eq.~\eqref{eq:soft_exp}~\cite{He:2014laa,Strominger:2014pwa}.
 
\begin{figure}[h]
\centering
\begin{tikzpicture}[scale=1.2]

  \draw[double, line width=0.45pt, double distance=1.2pt]
    (-0.55,0) -- (2.55,0);

  \node[source] (src) at (0,0) {};
  \node[below=0.18cm of src] {$\mathcal I^{ab}(\omega')$};
  \node[source] (ins) at (2,0) {};
  \node[below=0.18cm of ins] {$\mathcal I^{cd}(\omega-\omega')$};

  \node[vertex] (vtx) at (1,1.5) {};
  \node (ext) at (1,3.0) {};
  \node[above=0.10cm of ext] {$\sigma^*_{ij}(\omega,\mathbf{k})$};

  \draw[graviton] (src) -- (vtx)
    node[midway, above left] {\small $(\omega',\mathbf{q})$};
  \draw[graviton] (ins) -- (vtx)
    node[midway, above right] {\small $(\omega-\omega',\mathbf{k}-\mathbf{q})$};
  \draw[graviton] (vtx) -- (ext);
 
\end{tikzpicture}
\caption{Memory diagram. One-loop process with two retarded 
propagators. No static propagator --- the TT radiative amplitude is finite, with no divergence. 
The $1/\omega$ soft pole is the Weinberg soft graviton factor.}
\label{fig:Feyn_mem}
\end{figure}
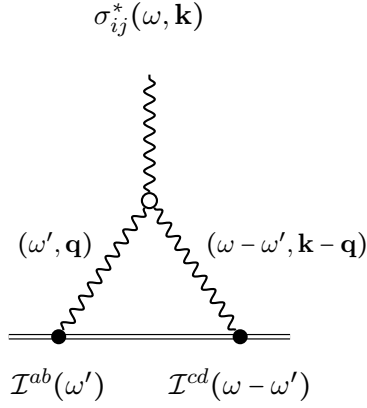

Fig.~\ref{fig:Feyn_mem} shows the diagram of the memory process, involving two mass-quadrupoles and an external graviton with polarization $\sigma_{ij}(\omega,\mathbf{k})$, with
$|\mathbf{k}|=\omega$.

 With
\begin{equation}
 D_1\equiv \mathbf q^2-(\omega'+i0^+)^2,\qquad
 D_2\equiv (\mathbf k-\mathbf q)^2-(\omega-\omega'+i0^+)^2\,,
 \label{eq:D12def}
\end{equation}
 the memory amplitude takes the form~\cite{Almeida:2024lbv}
\begin{equation}
 i\mathcal A^{\rm mem}_{ij}(\omega,\mathbf k)
 =\int\frac{d\omega'}{2\pi}\,
 \mathcal I^{ab}(\omega')\mathcal I^{cd}(\omega-\omega')
 \int_{\mathbf q}
 \frac{\mathcal N_{ij\,ab\,cd}
 (\mathbf q,\mathbf k;\omega',\omega-\omega')}{D_1D_2}.
 \label{eq:memory_NRGR_integrand}
\end{equation}
Here $\mathcal N_{ij}{}^{ab,cd}$ denotes the complete contraction of the two
source-vertex tensor structures, the radiative propagator numerators, and the
gauge-fixed $\sigma\sigma\sigma$ vertex.  Since the interaction vertex carries two
derivatives, the loop integral in \eqref{eq:memory_NRGR_integrand} reduces to the following structures:
\begin{equation}
 B_0\equiv \int_{\mathbf q}\frac{1}{D_1D_2},\qquad
 B_i\equiv \int_{\mathbf q}\frac{q_i}{D_1D_2},\qquad
 B_{ij}\equiv \int_{\mathbf q}\frac{q_iq_j}{D_1D_2}.
 \label{eq:memory_tensor_bubbles}
\end{equation}
The scalar master integral $B_0$ is computed in \eqref{eq:mem_master} where it is shown that it has a logarithmic divergence for $k\to \omega$. 
Writing $B_i\equiv k_iB_1$ and
$B_{ij}\equiv \delta_{ij}B_{00}+ \frac{k_i k_j}{k^2} B_{11}$, the identities
\begin{equation}
 2\mathbf k\cdot\mathbf q=k^2-(\omega-\omega')^2+\omega'^2+(D_1-D_2),
 \qquad
 \mathbf q^2=D_1+\omega'^2\
\end{equation}
give, for $d \rightarrow 3$ with $c\equiv (k^2+\omega_1^2-\omega_2^2)/2$ and
$T_a\equiv\int_{\mathbf q}D_a^{-1}=i\omega_a/(4\pi)$,
\begin{equation}
 \begin{aligned}
 B_i&=k_i\left[\frac{c}{k^2}B_0
 +\frac{T_2-T_1}{2k^2}\right],
 \\
 B_{00}
 &=
 \frac{1}{2}\left[\left(\omega_1^2-\frac{c^2}{k^2}\right)B_0
 +\frac{c}{2k^2}T_1
 +\frac{k^2-c}{2k^2}T_2 \right],
 \\
 B_{11}&= \frac 12\pa{\frac{3c^2}{k^2}-\omega_1^2}B_0-\frac{3c}{4k^2}T_1+\frac 14\pa{1+\frac{3c}{k^2}}T_2\,,
 \end{aligned}
 \label{eq:memory_tensor_reduction}
\end{equation}
where we used $\omega_1\equiv\omega'$ and
$\omega_2\equiv\omega-\omega'$, so that
$\omega=\omega_1+\omega_2$.

For the on-shell emitted graviton, $k=\omega$ and
$c=\omega\omega_1$, $B_0$ drops out of $B_{00}$, and
\begin{equation}
 \left.B_{00}\right|_{k=\omega}
 =\frac{\omega_1T_1+\omega_2T_2}{4\omega}
 =\frac{i}{16\pi\omega}\left(\omega_1^2+\omega_2^2\right)\,,
 \label{eq:memory_ranktwo_bubble_onshell}
\end{equation}
where $T_a$ are computed in Appendix \ref{sec:app_master}.
The remaining two coefficients become
\begin{equation}
 \left.B_i\right|_{k=\omega}
 =\hat n_i\pa{\omega_1 B_0+\frac{i}{8\pi\omega}(\omega_2-\omega_1)}\,,
 \qquad
 \left.B_{11}\right|_{k=\omega}
 =\omega_1^2 B_0
 +\frac{i}{16\pi\omega}
 \left[-3\omega_1^2+(4\omega_1+\omega_2)\omega_2\right]\,.
 \label{eq:memory_vector_bubble_onshell}
\end{equation}
The part of the integrals proportional to the $B_0$ master
integral happens to give a purely longitudinal contribution \cite{Almeida:2024lbv}
to the emitted waveform, hence is projected out of the radiative mode.\footnote{The contribution from all gravity polarizations in the internal propagators can add integrals with up to 6 free tensor indices, but still $B_0$ does not contribute to the radiative mode as shown in \cite{Almeida:2024lbv}.}
The $B_0$-independent finite parts in \eqref{eq:memory_ranktwo_bubble_onshell} and
 \eqref{eq:memory_vector_bubble_onshell} are all $\propto 1/\omega$, then the transverse part of the graph is the form
 \begin{equation}
  i\mathcal A^{\rm mem,TT}(\omega,\hat n)
  =\frac 1\omega\,
  \mathcal R^{\rm mem,TT}(\omega,\hat n)+O(\omega^0),
  \label{eq:memory_NRGR_onshell}
 \end{equation}
 for some $\mathcal R^{\rm mem,TT}$ which is finite as $\omega\to0$.

 Equation~\eqref{eq:memory_NRGR_onshell} identifies the pole with the frequency-space representation of the PN--MPM memory integral
 \eqref{eq:memory}. 
 
\subsubsection*{The Weinberg Soft Pole:}
 
In the limit $\omega\to 0$, $(\omega-\omega')\to-\omega'$:
\begin{equation}
    i\mathcal{A}_{\rm memory}\bigg|_{\omega\to 0} \sim
    \frac{iG}{\omega}\int\frac{d\omega'}{2\pi}\,\omega'^6\,
    |{\mathcal I}^{ij}(\omega')|^2 \sim
    \frac{iG\cdot\mathcal{E}_{\rm rad}}{\omega}
\end{equation}
where $\mathcal{E}_{\rm rad} =\frac G{5c^4}
\int\frac{d\omega'}{2\pi}\omega'^6|{\mathcal I}^{ij}(\omega')|^2$
is the total radiated energy (in the quadrupole approximation). This is the
Weinberg soft graviton pole $\sim 1/\omega$.
 
Key properties:
\begin{itemize}
    \item \textbf{Soft pole} $\sim 1/\omega$: Weinberg leading soft
    factor, classical limit of
    $S^{(0)}\sim p_a^\mu p_a^\nu\epsilon^*_{\mu\nu}/p_a\cdot k$
    \item \textbf{No $\log\omega$}: memory is finite, no hereditary
    logarithm
    \item \textbf{Zero-frequency displacement}: the $1/\omega$ pole
    in the amplitude gives a step-function-like displacement in the
    time domain --- the permanent memory effect
    \item \textbf{Coefficient}: proportional to total radiated energy,
    encoding the nonlinear sourcing of memory by gravitational wave
    stress-energy
\end{itemize}

\subsection{The Tail-of-Memory at 4PN}
\label{ssec:tom}
 
\begin{figure}[h]
\centering
\begin{tikzpicture}[scale=0.9]

  \draw[double, line width=0.45pt, double distance=1.2pt]
    (-0.55,0) -- (4.55,0);

  \node[source] (src) at (0,0) {};
  \node[below=0.18cm of src] {$\mathcal I^{ab}(\omega')$};
  \node[source] (ins) at (2,0) {};
  \node[below=0.18cm of ins] {$\mathcal I^{cd}(\omega-\omega')$};
  \node[source] (mass) at (4,0) {};
  \node[below=0.18cm of mass] {$M$};

  \node[vertex] (vtx1) at (1,1.5) {};
  \node[vertexstat] (vtx2) at (2.5,3.0) {};
  \node (ext) at (2.5,4.5) {};
  \node[above=0.10cm of ext] {$\sigma^*_{ij}(\omega,\mathbf{k})$};

  \draw[graviton] (src) -- (vtx1);
  \node[above left] at (0.35,0.85)
    {\scriptsize $(\omega',\mathbf{q})$};
  \draw[graviton] (ins) -- (vtx1);
  \node[above right] at (1.25,0.85)
    {\scriptsize $(\omega-\omega',\mathbf{p}-\mathbf{q})$};
  \draw[graviton] (vtx1) -- (vtx2);
  \node[above left] at (1.55,2.35) {\scriptsize $(\omega,\mathbf{p})$};
  \draw[static] (mass) -- (vtx2);
  \node[above right] at (3.25,1.45)
    {\scriptsize $(0,\mathbf{k}-\mathbf{p})$};
  \draw[graviton] (vtx2) -- (ext);
\end{tikzpicture}
\caption{Tail-of-memory diagram. Two-loop process with three retarded propagators 
(wavy lines) and one static propagator (dashed).}
\label{fig:tail_of_mem}
\end{figure}
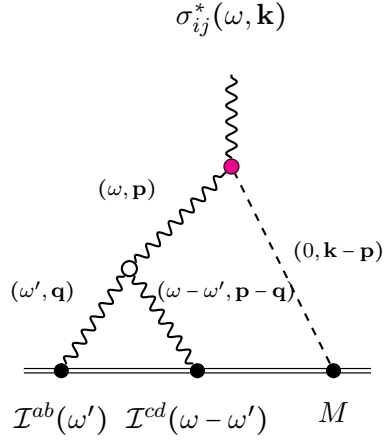

The tail-of-memory arises when the nonlinear memory itself undergoes tail scattering off the static background curvature generated by the total mass $M$. At the diagrammatic level, this corresponds to appending one $\sigma\sigma\phi$ bulk vertex and one static propagator to the memory diagram, promoting a one-loop process into a two-loop one. The crucial new element is the static propagator (dashed line in Fig.~\ref{fig:tail_of_mem}), which carries zero frequency. As we show below, it is precisely this propagator that generates the single IR pole $1/\epsilon_{\rm IR}$ and the associated physical $\log\omega$, yielding the sub-leading soft behavior $\sim\log\omega$ in the soft limit. This is the PN realization of the Sahoo--Sen coefficient $\tilde{h}_{\mu\nu}^{(\log)}$ in the soft expansion~\eqref{eq:soft_exp}~\cite{Sahoo:2018lxl,Saha:2019tub}.
 
The diagram involves two mass quadrupoles, a mass insertion, and two bulk
vertices.  The two quadrupole lines first meet at the $\sigma\sigma\sigma$
vertex $V_1$, producing an intermediate radiative line of momentum
$p^\mu=(\omega,\mathbf p)$.  The latter scatters onto the static mass
insertion through the $\sigma\sigma\phi$ vertex $V_2$. We denote the complete TT contraction of $V_1$ with $\sigma$ fields propagators sourced by ${\mathcal I}^{ab},{\mathcal I}^{cd}$, with
$\mathcal V_{1\,ab\,cd\,mn}(\omega_1,\mathbf q;
\omega_2,\mathbf p-\mathbf q)$.  Its explicit KK rule follows from the
cubic bulk action \cite{Foffa:2013qca,Almeida:2024lbv}.
 
The vertex $V_2$ connects one internal radiative mode, the static mode, and
the external $\sigma^*_{ij}$.  It involves one time derivative for each
$\sigma$ (see e.g.~eq.~(13) of \cite{Foffa:2013qca}), and its contraction
with a $\sigma$ propagator $P_{\sigma\sigma}$ gives
\begin{equation}
   P_{\sigma_{kl}\sigma_{ab}} V^{(3)}_{2\ abij} = -2i\frac{\omega^2}{\Lambda} g^{i(k}g^{l)j}\,.
   \label{eq:PssV2}
\end{equation}
 
The restriction to the $\sigma\sigma\phi$ sector is justified because it is the
only sector generating the IR poles relevant for the soft analysis; see
Appendix~\ref{sec:app_polarsector}.  With
\begin{equation}
 D_1\equiv\mathbf q^2-(\omega_1+i0^+)^2,\qquad
 D_3\equiv(\mathbf p-\mathbf q)^2-(\omega_2+i0^+)^2,\qquad
 D_p\equiv\mathbf p^2-(\omega+i0^+)^2,
 \label{eq:tom_denominators}
\end{equation}
where $\omega_1\equiv\omega'$ and $\omega_2\equiv\omega-\omega'$, the
$\mathbf{q}$-subgraph, including both source vertices and the complete $V_1$
contraction, is the off-shell continuation of the memory graph:
\begin{align}
 \mathcal H^{\rm mem}_{kl}(\omega,\mathbf p)
 \equiv{}&
 \int\frac{d\omega'}{2\pi}\,
 \omega_1^2\omega_2^2\,
 {\mathcal I}^{ab}(\omega_1){\mathcal I}^{cd}(\omega_2)
 \int_{\mathbf q}
 \frac{\mathcal V_{1\,ab\,cd\,kl}
 (\omega_1,\mathbf q;\omega_2,\mathbf p-\mathbf q)}
 {D_1D_3}.
 \label{eq:tom_memory_subgraph}
\end{align}
The source-vertex factors and the normalization of $V_1$ are included in
$\mathcal H^{\rm mem}$.  The only difference from the memory graph of
section~\ref{sec:NRGR_memory} is that the outgoing line has spatial momentum
$\mathbf p$, and is consequently off shell before the final $\mathbf p$-integral is
performed.

The two-loop amplitude can therefore be rearranged, without approximations,
as a tail kernel acting on this subgraph:
\begin{equation}
 \left.i\mathcal A_{\rm tom}^{\rm TT}\right|_{\text{IR}}
 =i2 \frac{M\omega^2}{\Lambda^2}\,\sigma^{*}_{ij}(\omega,\mathbf k)
 \Pi_{ijkl}(\hat n)\int_{\mathbf p}
 \frac{\mathcal H^{\rm mem}_{kl}(\omega,\mathbf p)}
 {D_p(\mathbf k-\mathbf p)^2}.
 \label{eq:tom_memory_factorization}
\end{equation}
The factor $M\omega^2$ follows from the static source and the projected
$\sigma\sigma\phi$ vertex displayed in \eqref{eq:PssV2}.  Eq.~\eqref{eq:tom_memory_factorization} contains all the four propagators: $D_1$
and $D_3$ belong to the memory subgraph, whereas $D_p$ and
$(\mathbf k-\mathbf p)^2$ form the tail loop.

To isolate the non-analytic part, we substitute $\mathbf p\equiv\mathbf k-\mathbf r$
and impose $\mathbf k^2=\omega^2$.  The two tail denominators are then
\begin{equation}
 D_p=-2\mathbf k\!\cdot\!\mathbf r+\mathbf r^2,
 \qquad (\mathbf k-\mathbf p)^2=\mathbf r^2\,,
 \label{eq:tom_pinch_denominators}
\end{equation}
which vanish simultaneously for $\mathbf r=0$. Expanding the numerator of the integrand of (\ref{eq:tom_memory_factorization}) around ${\mathbf r}=0$, we obtain
\begin{equation}
 \Pi_{ijkl}\mathcal H^{\rm mem}_{kl}
 (\omega,\mathbf k-\mathbf r)
 =\mathcal H^{\rm TT,mem}_{ij}(\omega,\mathbf k)+O(r).
 \label{eq:tom_memory_pinch_expansion}
\end{equation}
At $\mathbf p=\mathbf k$, both $D_3$ and $V_1$ in
\eqref{eq:tom_memory_subgraph} are those of the ordinary on-shell memory
graph.  Hence
\begin{equation}
 \sigma^*_{ij}\mathcal H_{ij}^{\rm TT,mem}(\omega,\mathbf k)
 =i\mathcal A_{\rm mem}^{\rm TT}(\omega,\hat n)\,,
 \label{eq:tom_memory_onshell}
\end{equation}
which can be plugged into \eqref{eq:tom_memory_factorization} together with \eqref{eq:tom_pinch_denominators} to obtain
\begin{equation}
 \left.i\mathcal A_{\rm tom}^{\rm TT}\right|_{\rm IR}
=2 i\, \frac{M\omega^2}{\Lambda^2}\,\mathcal A^{\rm mem, TT}(\omega,\hat n)
 \underbrace{\int_{\mathbf r}
 \frac{1}{\pa{-2\mathbf k\!\cdot\!\mathbf r+\mathbf r^2}\mathbf{r}^2}}_{I_1(\omega)}
 +\frac{O(r)}{\pa{-2\mathbf k\!\cdot\!\mathbf r+\mathbf r^2}\mathbf{r}^2}\,,
 \label{eq:tom_memory_I1}
\end{equation}
where the second term in the integral is clearly not IR divergent.
Here
\begin{align}
   I_1(\omega)\equiv &\left.\int_\mathbf{p}\frac{1}{[\mathbf{p}^2-(\omega+i0^+)^2]
    \cdot(\mathbf{k}-\mathbf{p})^2}\right|_{\mathbf{k}=\omega\hat n} \\
    =& -\frac 1{8\pi\omega}\left[\frac{1}{\epsilon_{\rm IR}}
    - \frac{1}{2}\log\frac{\omega^2}{\tilde\mu^2}
    + \frac{i\pi}{2}{\rm sgn}(\omega) + O(\epsilon)\right]\equiv -\frac 1{8\pi\omega}\paq{\mathcal{L}(\omega)+O(\epsilon)}\,,
\end{align}
and
\begin{equation}
\mathcal{L}(\omega)\equiv
\frac{1}{\epsilon_{\rm IR}}
-\frac{1}{2}\log\frac{\omega^2}{\tilde\mu^2}
+\frac{i\pi}{2}{\rm sgn}(\omega)\,.
\label{eq:L_definition}
\end{equation}
see Appendix \ref{sec:app_master} for details.  Combining
\eqref{eq:tom_memory_I1} with the on-shell memory result
\eqref{eq:memory_NRGR_onshell} gives
\begin{equation}
 \left.\mathcal A_{\rm tom}^{\rm TT}\right|_{\rm IR}
 \sim (iGM\omega)\,\mathcal A_{\rm mem}^{\rm TT}(\omega,\hat n)
 \mathcal L(\omega).
 \label{eq:tom_soft_factorization}
\end{equation}
Thus the $\omega^2$ supplied by $V_2$ is cancelled by one factor
$1/\omega$ from the on-shell memory amplitude and one from
$I_1(\omega)$.
 
\subsubsection*{Divergence Structure:}
 
It is worthwhile to summarize the divergence structure of the amplitude and its soft behavior. We have the following contributions:
\begin{itemize}
    \item \textbf{Single IR pole} $1/\epsilon_{\rm IR}$: from the loop involving the  static propagator resulting into $I_1(\omega)$;
    \item \textbf{$\log(\omega^2/\tilde\mu^2)$}: physical hereditary log;
    \item \textbf{$i\pi\,{\rm sgn}(\omega)$}: it is real relative to the leading order emission term, thus contributing to the flux $(\sim \omega^2|{\mathcal A}^2|)$ at $GM\omega\sim v^3$ order with respect to the leading one.
\end{itemize}
 
\subsubsection*{Soft Behavior at $\omega\to 0$:}
\label{sec:tom_soft}
 
At $\omega\to 0$, the STF source tensor $\tilde{\mathcal{Q}}_{ij}(\omega)$ tends to
a finite tensor $\tilde{\mathcal{Q}}_{ij}(0)$. Therefore:
\begin{equation}
    \left.i\mathcal{A}^{\rm TT}_{\rm tom}\right|_{\rm IR} \sim
    GM\cdot\sigma^{*ij}\tilde{\mathcal{Q}}_{ij}(0)\cdot\log\omega\,.
\end{equation}
 
The amplitude has pure $\log\omega$ behavior in the soft limit.
Note that the source frequencies $\omega'$ are independent of the
external soft $\omega$ --- the dominant contribution to the $\omega'$
integral comes from $\omega'\sim\omega_{\rm orb}$, a fixed physical
scale set by the orbital dynamics.

\section{The tail-of-tail-of-memory and its higher generalization to $n$-tail insertions} \label{sec:higher}
 
Having established the pattern at one and two loops, we now extend the computation to higher orders. The key observation from the tail-of-memory analysis is that the IR structure is entirely controlled by the number of static propagators: each mass insertion $M$ contributes one static propagator, and each of them evaluates to the same master integral $I_1(\omega)$ since all carry the same external frequency $\omega$. This additive accumulation of $I_1(\omega)$ factors is the microscopic origin of the IR exponentiation that we establish in section~\ref{sec:IR_softexp}.
 
We begin by computing the tail-of-tail-of-memory explicitly at three loops (two mass insertions, 5.5PN order), then derive the general amplitude for $n$ mass insertions.
 
\subsection{The tail-of-tail-of-memory}
 
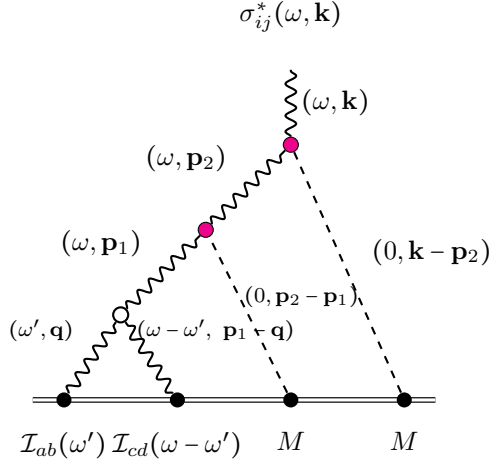
\begin{figure}[h]
\centering
\begin{tikzpicture}[scale=0.75]

  \draw[double, line width=0.45pt, double distance=1.2pt]
    (-0.55,0) -- (6.55,0);

  \node[source] (src) at (0,0) {};
  \node[below=0.18cm of src] {\small $\mathcal I_{ab}(\omega')$};
  \node[source] (ins) at (2,0) {};
  \node[below=0.18cm of ins] {\small $\mathcal I_{cd}(\omega-\omega')$};
  \node[source] (mass1) at (4,0) {};
  \node[below=0.18cm of mass1] {\small $M$};
  \node[source] (mass2) at (6,0) {};
  \node[below=0.18cm of mass2] {\small $M$};

  \node[vertex] (vtx1) at (1,1.5) {};
  \node[vertexstat] (vtx2) at (2.5,3.0) {};
  \node[vertexstat] (vtx3) at (4.0,4.5) {};
  \node (ext) at (4.0,6.0) {};
  \node[above=0.10cm of ext]
    {\small $\sigma^*_{ij}(\omega,\mathbf{k})$};

  \draw[graviton] (src) -- (vtx1);
  \node[above left] at (0.35,0.85)
    {\scriptsize $(\omega',\mathbf{q})$};
  \draw[graviton] (ins) -- (vtx1);
  \node[above right] at (1.1,0.85)
    {\scriptsize $(\omega-\omega', ~\mathbf{p}_1-\mathbf{q})$};
  \draw[graviton] (vtx1) -- (vtx2);
  \node[above left] at (1.55,2.35) {\small $(\omega,\mathbf{p}_1)$};
  \draw[static] (mass1) -- (vtx2);
  \node[above right] at (3,1.45)
    {\scriptsize $(0,\mathbf{p}_2-\mathbf{p}_1)$};
  \draw[graviton] (vtx2) -- (vtx3);
  \node[above left] at (3.05,3.85) {\small $(\omega,\mathbf{p}_2)$};
  \draw[static] (mass2) -- (vtx3);
  \node[above right] at (5.25,2.2)
    {\small $(0,\mathbf{k}-\mathbf{p}_2)$};
  \draw[graviton] (vtx3) -- (ext);
  \node[right] at (4,5.25) {\small $(\omega,\mathbf{k})$};

\end{tikzpicture}
\caption{Tail-of-tail-of-memory diagram. Three-loop process with
four retarded propagators (wavy lines) and two
static propagators (dashed). The finite memory subgraph
connects the two source quadrupoles. }
\end{figure}
 
The tail-of-tail-of-memory is a three-loop process obtained by adding
a second mass insertion to the tail-of-memory diagram. The fields sourced by the 
quadrupoles $\mathcal I_{ab}(\omega-\omega')$ and $\mathcal I_{cd}(\omega')$ meet at the
same $\sigma\sigma\sigma$ vertex $V^{(3)}_1$, while the two subsequent
$\sigma\sigma\phi$ vertices generate nested tail kernels.
 
The three-loop amplitude may therefore be written directly in
terms of the complete off-shell memory form factor
$\mathcal H^{\rm mem}_{kl}$ of \eqref{eq:tom_memory_subgraph}:
\begin{equation}
 \left.i\mathcal A_{\rm ttom}^{\rm TT}\right|_{\text{IR}}
 =\left(i2\frac{M\omega^2}{\Lambda^2}\right)^2
 \sigma^*_{ij}(\omega,\mathbf k)\Pi_{ijkl}(\hat n)
 \int_{\mathbf p_1,\mathbf p_2}
 \frac{\mathcal H^{\rm mem}_{kl}(\omega,\mathbf p_1)}
 {D_{p_1}(\mathbf p_2-\mathbf p_1)^2D_{p_2}
  (\mathbf k-\mathbf p_2)^2}\,.
 \label{eq:ttom_memory_factorization}
\end{equation}
where $D_{p_a}\equiv\mathbf p_a^2-(\omega+i0)^2$ for $a=1,2$.
Thus the two source propagators and the full $V_1$ contraction are kept
inside $\mathcal H^{\rm mem}_{kl}$.

To isolate the leading non-analytic part, we set
\begin{equation}
 \mathbf p_2=\mathbf k-\mathbf r_2,
 \qquad
 \mathbf p_1=\mathbf k-\mathbf r_2-\mathbf r_1,
 \qquad \mathbf k^2=\omega^2.
 \label{eq:ttom_pinch_variables}
\end{equation}
All four tail denominators in \eqref{eq:ttom_memory_factorization} are then
\begin{equation}
 \begin{aligned}
 \mathbf p_2^2-(\omega+i0)^2
 &=-2\mathbf k\!\cdot\!\mathbf r_2+\mathbf r_2^2,
 & (\mathbf k-\mathbf p_2)^2&=\mathbf r_2^2,\\
 \mathbf p_1^2-(\omega+i0)^2
 &=-2\mathbf k\!\cdot\!(\mathbf r_1+\mathbf r_2)
   +(\mathbf r_1+\mathbf r_2)^2,
 & (\mathbf p_2-\mathbf p_1)^2&=\mathbf r_1^2.
 \end{aligned}
 \label{eq:ttom_pinch_denominators}
\end{equation}
Thus the tail-of-tail-of-memory amplitude has the form
\begin{equation}
 \begin{aligned}
 \left.i\mathcal A_{\rm ttom}^{\rm TT}\right|_{\text{IR}}
 ={}&\left(i2\frac{M\omega^2}{\Lambda^2}\right)^2
 \sigma^*_{ij}\Pi_{ijkl}(\hat n)
 \int_{\mathbf r_1,\mathbf r_2}
 \frac{\mathcal H^{\rm mem}_{kl}
 (\omega,\mathbf k-\mathbf r_1-\mathbf r_2)}
 {\mathbf r_1^2\mathbf r_2^2
  \bigl[-2\mathbf k\!\cdot\!(\mathbf r_1+\mathbf r_2)
  +(\mathbf r_1+\mathbf r_2)^2\bigr]
  \bigl[-2\mathbf k\!\cdot\!\mathbf r_2+\mathbf r_2^2\bigr]}\,.
 \end{aligned}
 \label{eq:ttom_coupled_pinch}
\end{equation}
Expanding the numerator of the integrand around $\mathbf{r}_i = 0$ gives
\begin{equation}
 \sigma^*_{ij}\Pi_{ijkl}(\hat n)
 \mathcal H^{\rm mem}_{kl}
 (\omega,\mathbf k-\mathbf r_1-\mathbf r_2)
 =i\mathcal A_{\rm mem}^{\rm TT}(\omega,\hat n)
 +O(\mathbf r_1,\mathbf r_2),
 \label{eq:ttom_memory_pinch_expansion}
\end{equation}

To isolate the leading IR behavior, we can approximate the integration over $\mathbf{r}_2$ as
\begin{align*}
&\int_{\mathbf{r}_2}\frac{1}{\paq{-2\mathbf k\!\cdot\!(\mathbf r_1+\mathbf r_2)
  +(\mathbf r_1+\mathbf r_2)^2}
  \paq{-2\mathbf k\!\cdot\!\mathbf r_2
  +\mathbf r_2^2}{\mathbf r}_2^2}\\
  &\sim \frac{1}{\paq{-2\mathbf k\!\cdot\!\mathbf r_1
   +\mathbf r_1^2}}\int_{\mathbf{r}_2}
   \frac{1+O(r_2)}{\paq{-2\mathbf k\!\cdot\!\mathbf r_2 +\mathbf r_2^2}\mathbf{r}_2^2}\,,
\end{align*}
thus reproducing the IR divergence for $\mathbf{r}_2\to 0$, and the integration over $\mathbf{r}_2$ returns $I_1(\omega)$ at leading soft behavior.
The leading soft behavior of the integration over ${\mathbf r}_1$
is then obtained in analogy with the tail-of-memory as
done in section~\ref{ssec:tom}.

Equation \eqref{eq:ttom_coupled_pinch} is written as
\begin{align}
 \left.i\mathcal A_{\rm ttom}^{\rm TT}\right|_{\text{IR}}
 ={}&
 \left(i2\frac{M\omega^2}{\Lambda^2}\right)^2
 \Bigg\{
 i\mathcal A_{\rm mem}^{\rm TT}(\omega,\hat n)
 \underbrace{\int_{\mathbf r_1}
 \frac{1}{\bigl(-2\mathbf k\!\cdot\!\mathbf r_1+\mathbf r_1^2\bigr)
 \mathbf r_1^2}}_{I_1(\omega)} \underbrace{\int_{\mathbf r_2}
 \frac{1}{\bigl(-2\mathbf k\!\cdot\!\mathbf r_2+\mathbf r_2^2\bigr)
 \mathbf r_2^2}}_{I_1(\omega)} \\
 &\quad+\int_{\mathbf r_1,\mathbf r_2}
 \frac{O(\mathbf r_1,\mathbf r_2)}
 {\mathbf r_1^2\mathbf r_2^2
  \bigl[-2\mathbf k\!\cdot\!(\mathbf r_1+\mathbf r_2)
        +(\mathbf r_1+\mathbf r_2)^2\bigr]
  \bigl[-2\mathbf k\!\cdot\!\mathbf r_2+\mathbf r_2^2\bigr]}
 \Bigg\}.
 \label{eq:ttom_I1_iteration}
\end{align}

Thus only the product of two $I_1(\omega)$ integrals contributes
to the leading non-analytic part, i.e., the double logarithm.
Using
\eqref{eq:memory_NRGR_onshell} and \eqref{eq:master_integral_tail} in
\eqref{eq:ttom_I1_iteration} gives
\begin{equation}
 \left.\mathcal A_{\rm ttom}^{\rm TT}\right|_{\rm IR}
 \sim (iGM\omega)^2\,
 \mathcal A_{\rm mem}^{\rm TT}(\omega,\hat n)
 [\mathcal L(\omega)]^2.
 \label{eq:ttom_soft_factorization}
\end{equation}
Since $\mathcal A_{\rm mem}^{\rm TT}\sim\omega^{-1}$, this is
proportional to $(GM)^2\omega[\mathcal L(\omega)]^2$, as required.
 
Writing the source dependence in terms of
$\tilde{\mathcal Q}_{ij}(\omega)$, the same result is
\begin{align}
    i\mathcal{A}_{\rm ttom} \simeq i(GM)^2\cdot\omega
    \cdot[\mathcal{L}(\omega)]^2\cdot
    \sigma^{*ij}\tilde{\mathcal{Q}}_{ij}(\omega) \,.
    \label{eq:ttom_result}
\end{align}
 
\subsection*{Divergence Structure:}
The divergence structure follows directly from expanding $[\mathcal{L}(\omega)]^2$:
\begin{itemize}
    \item \textbf{Double IR pole} $1/\epsilon^2_{\rm IR}$: from two loop integrations involving the static propagators resulting in $[I_1(\omega)]^2$
    \item \textbf{Cross terms} $1/\epsilon_{\rm IR}\left(\log\omega+i\pi\right)$: from binomial expansion
    \item \textbf{$(\log\omega)^2$}: physical double hereditary logarithm
\end{itemize}
Other polarization sectors are addressed in Appendix~\ref{sec:app_polarsector}.
\subsection*{Soft Behavior at $\omega\to 0$:}
 
After regularizing the IR divergences, see section~\ref{sec:IR_softexp}, the dominant term at $\omega\to 0$ comes from the $\log^2\omega$ piece of $[\mathcal{L}(\omega)]^2$. Combined with the explicit $\omega$ prefactor and the finite limit of $\mathcal{Q}_{ij}(\omega)$:
\begin{equation}
    i\mathcal{A}_{\rm ttom}\bigg|_{\omega\to 0}
    \sim i(GM)^2\cdot\sigma^{*ij}\mathcal{Q}_{ij}(0)
    \cdot\omega\,(\log\omega)^2 \,.
\end{equation}
The comparison table with the tail-of-memory makes the pattern manifest:
\begin{center}
\begin{tabular}{|l|c|c|}
\hline
\textbf{Property} &
\textbf{Tail-of-memory} &
\textbf{Tail-of-tail-of-memory}\\
\hline
Mass insertions & 1 & 2\\
IR poles & $1/\epsilon_{\rm IR}$ & $1/\epsilon^2_{\rm IR}$\\
Loop structure &
$I_1(\omega)$ &
$[I_1(\omega)]^2$\\
Net external $\omega$ & $\omega^0$ & $\omega^1$\\
Soft behavior & $\log\omega$ & $\omega(\log{\omega})^2$\\
PN order & 4PN & 5.5PN\\
\hline
\end{tabular}
\end{center}
 
The pattern is clear: each additional mass insertion $M$ adds one
$\sigma\sigma\phi$ vertex, one $I_1(\omega)$, and one power each of
$GM$, $\log\omega$, and external $\omega$.  The finite prefactor is fixed
by the complete vertex contraction.  The general rule for
$n$ tail insertions is:
 
\begin{equation}
    i\mathcal{A}_{n\text{ tails}}\bigg|_{\omega\to 0}
    \sim (GM)^n\cdot\sigma^{*ij}\mathcal{Q}_{ij}(0)
    \cdot\omega^{n-1}\log^n\omega
\end{equation}
 
\subsection{Generalized Computation: Memory with $n$-tail Insertions}
 
We now extend the nested-tail analysis of section~4.1 to $n$ mass insertions.
The graph contains the finite memory subgraph, $n$ $\sigma\sigma\phi$
vertices, and $n$ static propagators.  The emitted graviton carries
$(\omega,\mathbf k)$, with $\mathbf k^2=\omega^2$.
 
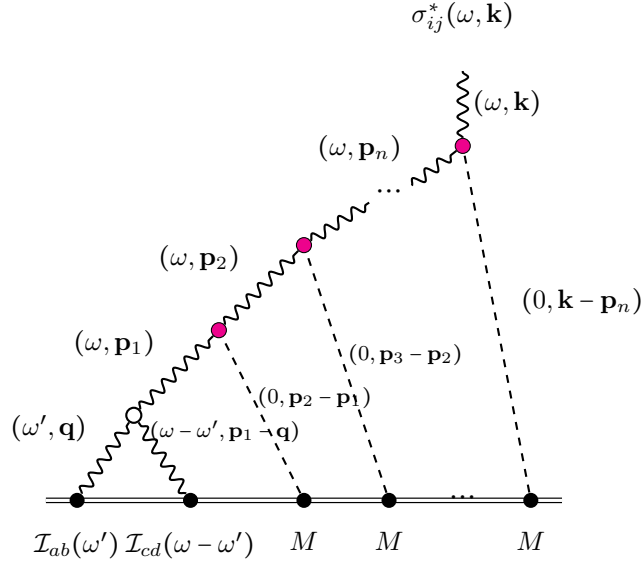
\begin{figure}[h]
\centering
\begin{tikzpicture}[scale=0.75]

  \draw[double, line width=0.45pt, double distance=1.2pt]
    (-0.55,0) -- (8.55,0);

  \node[source] (src) at (0,0) {};
  \node[below=0.18cm of src] {\small $\mathcal I_{ab}(\omega')$};
  \node[source] (ins) at (2,0) {};
  \node[below=0.18cm of ins] {\small $\mathcal I_{cd}(\omega-\omega')$};
  \node[source] (mass1) at (4,0) {};
  \node[below=0.18cm of mass1] {\small $M$};
  \node[source] (mass2) at (5.5,0) {};
  \node[below=0.18cm of mass2] {\small $M$};
  \node at (6.8,0) {\large $\cdots$};
  \node[source] (massn) at (8,0) {};
  \node[below=0.18cm of massn] {\small $M$};

  \node[vertex] (vtx1) at (1,1.5) {};
  \node[vertexstat] (vtx2) at (2.5,3.0) {};
  \node[vertexstat] (vtx3) at (4.0,4.5) {};
  \node at (5.5,5.4) {\large $\cdots$};
  \node[vertexstat] (vtxn) at (6.8,6.25) {};
  \node (ext) at (6.8,7.75) {};
  \node[above=0.10cm of ext]
    {\small $\sigma^*_{ij}(\omega,\mathbf{k})$};

  \draw[graviton] (src) -- (vtx1);
  \node[above left] at (0.35,0.85)
    {\small $(\omega',\mathbf{q})$};
  \draw[graviton] (ins) -- (vtx1);
  \node[above right] at (1.1,0.85)
    {\scriptsize $(\omega-\omega',\mathbf{p}_1-\mathbf{q})$};
  \draw[graviton] (vtx1) -- (vtx2);
  \node[above left] at (1.55,2.35) {\small $(\omega,\mathbf{p}_1)$};
  \draw[static] (mass1) -- (vtx2);
  \node[above right] at (3,1.45)
    {\scriptsize $(0,\mathbf{p}_2-\mathbf{p}_1)$};
  \draw[graviton] (vtx2) -- (vtx3);
  \node[above left] at (3.05,3.85) {\small $(\omega,\mathbf{p}_2)$};
  \draw[static] (mass2) -- (vtx3);
  \node[above right] at (4.6,2.2)
    {\scriptsize $(0,\mathbf{p}_3-\mathbf{p}_2)$};
  \draw[graviton] (vtx3) -- (5.15,5.25);
  \draw[graviton] (5.9,5.55) -- (vtxn);
  \node[above left] at (5.9,5.8) {\small $(\omega,\mathbf{p}_n)$};
  \draw[static] (massn) -- (vtxn);
  \node[above right] at (7.65,3.1)
    {\small $(0,\mathbf{k}-\mathbf{p}_n)$};
  \draw[graviton] (vtxn) -- (ext);
  \node[right] at (6.8,7.0) {\small $(\omega,\mathbf{k})$};

\end{tikzpicture}
\caption{General $n$-tail-of-memory diagram. The process has
$n+2$ retarded propagators and $n$ static propagators (dashed).
The finite memory subgraph connects the two source quadrupoles.}
\end{figure}
 
Keeping the finite memory subgraph off-shell, the exact $n$-tail-of-memory amplitude
can be organized as
\begin{align}
 i\mathcal A_{t^{n}\text{om}}^{\rm TT}
 ={}&\left(i2\frac{M\omega^2}{\Lambda^2}\right)^n
 \sigma^*_{ij}\Pi_{ijkl}(\hat n)
 \int_{\mathbf p_1,\ldots,\mathbf p_n}
 \frac{\mathcal H^{\rm mem}_{kl}(\omega,\mathbf p_1)}
 {\displaystyle
  \prod_{a=1}^{n}\bigl[\mathbf p_a^2-(\omega+i0)^2\bigr]\,
  \prod_{a=1}^{n-1}(\mathbf p_{a+1}-\mathbf p_a)^2\,
  (\mathbf k-\mathbf p_n)^2}\,.
 \label{eq:ntail_memory_factorization}
\end{align}
This is the direct $n$-tail extension of
\eqref{eq:ttom_memory_factorization}: the two source propagators and the
complete $V_1$ contraction remain inside $\mathcal H^{\rm mem}_{kl}$.

To expose the leading non-analytic terms, we introduce
\begin{equation}
 \mathbf p_a=\mathbf k-\mathbf R_a,\qquad
 \mathbf R_a\equiv\sum_{b=a}^{n}\mathbf r_b,\qquad a=1,\ldots,n.
 \label{eq:ntail_pinch_variables}
\end{equation}
In particular,
\begin{equation}
 \mathbf p_1=\mathbf k-\sum_{a=1}^{n}\mathbf r_a ,\qquad \mathbf p_2=\mathbf k-\sum_{a=2}^{n}\mathbf r_a , \qquad \dots \qquad \mathbf{p}_n= \mathbf k-\mathbf{r}_n\,.
 \label{eq:ntail_p1}
\end{equation}
The retarded and static denominators in
\eqref{eq:ntail_memory_factorization} become
\begin{equation}
 \mathbf p_a^2-(\omega+i0)^2
 =-2\mathbf k\!\cdot\!\mathbf R_a+\mathbf R_a^2,\qquad
 (\mathbf p_{a+1}-\mathbf p_a)^2=\mathbf r_a^2\quad(a=1,\ldots,n-1),\qquad
 (\mathbf k-\mathbf p_n)^2=\mathbf r_n^2.
 \label{eq:ntail_pinch_denominators}
\end{equation}
Hence, the $n$-tail-of-memory amplitude can be written as
\begin{align}
 i\mathcal A_{t^{n}\text{om}}^{\rm TT}
 ={}&\left(i2\frac{M\omega^2}{\Lambda^2}\right)^n
 \sigma^*_{ij}\Pi_{ijkl}(\hat n)
 \int_{\mathbf r_1,\ldots,\mathbf r_n}
 \frac{\mathcal H^{\rm mem}_{kl}(\omega,\mathbf k-\mathbf R_1)}
 {\displaystyle\prod_{a=1}^{n}\mathbf r_a^2
  \bigl[-2\mathbf k\!\cdot\!\mathbf R_a+\mathbf R_a^2\bigr]}.
 \label{eq:ntail_coupled_pinch}
\end{align}
Expanding the numerator of the integrand around $\mathbf{r}_a = 0$ as before,
\begin{equation}
 \sigma^*_{ij}\Pi_{ijkl}(\hat n)
 \mathcal H^{\rm mem}_{kl}(\omega,\mathbf k-\mathbf R_1)
 \simeq i\mathcal A_{\rm mem}^{\rm TT}(\omega,\hat n)
 +O(\mathbf r_1,\ldots,\mathbf r_n).
 \label{eq:ntail_memory_pinch_expansion}
\end{equation}

Iterating the previous construction of the tail-of-memory and tail-of-tail-of-memory amplitude, one can find
\begin{align}
 i\mathcal A_{t^{n}\text{om}}^{\rm TT}
 ={}&\left(i2\frac{M\omega^2}{\Lambda^2}\right)^n\frac 1{n!}
 \Bigg\{
 i\mathcal A_{\rm mem}^{\rm TT}(\omega,\hat n)
 \prod_{a=1}^{n}
 \underbrace{\int_{\mathbf r_a}
 \frac{1}{\mathbf r_a^2
 \bigl(-2\mathbf k\!\cdot\!\mathbf r_a+\mathbf r_a^2\bigr)}}_{I_1(\omega)}
+\int_{\mathbf r_1,\ldots,\mathbf r_n}
 \frac{O(\mathbf r_1,\ldots,\mathbf r_n)}
 {\displaystyle\prod_{a=1}^{n}\mathbf r_a^2
  \bigl[-2\mathbf k\!\cdot\!\mathbf R_a+\mathbf R_a^2\bigr]}
 \Bigg\}.
 \label{eq:ntail_I1_iteration_bis}
\end{align}

Only the product of the $n$ displayed master
integrals contributes to the leading $(\log\omega)^n$ term. The remaining terms are subleading in the soft non-analytic hierarchy. Using
\eqref{eq:master_integral_tail},
\begin{equation}
 \left.\mathcal A_{n\text{-tail}}^{\rm TT}\right|_{\rm IR}
 \sim (iGM\omega)^n\,
 \mathcal A_{\rm mem}^{\rm TT}(\omega,\hat n)
 [\mathcal L(\omega)]^n.
 \label{eq:ntail_soft_factorization}
\end{equation}

The structure is clear: the same contracted STF source tensor multiplied by $(GM)^n$, an explicit $\omega^{n-1}$, and the $n$-fold hereditary logarithm $[\mathcal{L}(\omega)]^n$.
 
\subsection*{Divergence structure and soft behavior}
 
The divergence and soft structure follow directly from eq.~\eqref{eq:ntail_soft_factorization}:
\begin{itemize}
    \item \textbf{$n$-fold IR pole} $1/\epsilon^n_{\rm IR}$ from $[I_1(\omega)]^n$; lower poles $1/\epsilon^{n-j}_{\rm IR}\times\log^j\omega$ ($j=1,\ldots,n-1$) from the cross terms.
    \item \textbf{Dominant soft log} $\log^n\omega$ (physical, $j=n$ term); finite terms.
    \item \textbf{Net power} $\omega^{n-1}$: fixed by PN counting and by
    matching the finite full-vertex coefficient to the radiative multipole.
\end{itemize}
 
At $\omega\to 0$, $\mathcal{Q}_{ij}(\omega)$ has a finite limit independent of $n$, giving the universal soft behavior:
\begin{equation}
    i\mathcal{A}_n\bigg|_{\omega\to 0}
    \sim i(GM)^n\cdot\sigma^{*ij}\mathcal{Q}_{ij}(0)
    \cdot\omega^{n-1}(\log\omega)^n \,.
    \label{eq:soft_ntail}
\end{equation}
 
The soft behavior follows a universal pattern: 
 
\begin{center}
\begin{tabular}{|l|c|c|c|c|c|}
\hline
\textbf{Process} & $n$ &
\textbf{Static props} &
\textbf{IR poles} &
\textbf{Net $\omega$} &
\textbf{Soft behavior}\\
\hline
Memory & 0 & 0 & none &
$\omega^{-1}$ & $1/\omega$\\
Tail-of-memory & 1 & 1 &
$1/\epsilon_{\rm IR}$ &
$\omega^0$ & $\log\omega$\\
Tail-of-tail-of-memory & 2 & 2 &
$1/\epsilon^2_{\rm IR}$ &
$\omega^1$ & $\omega(\log\omega)^2$\\
Tail-of-tail-of-tail-of-memory & 3 & 3 &
$1/\epsilon^3_{\rm IR}$ &
$\omega^2$ & $\omega^2(\log\omega)^3$\\
$(n)$-tail-of-memory & $n$ & $n$ &
$1/\epsilon^n_{\rm IR}$ &
$\omega^{n-1}$ & $\omega^{n-1}(\log\omega)^n$\\
\hline
\end{tabular}
\end{center}
 
The general soft behavior $\omega^{n-1}\log^n\omega$ matches
the classical soft theorem prediction at sub$^{(n)}$-leading
order in the soft expansion, where each power of $\log\omega$
encodes one level of hereditary (tail) scattering off the background
mass $M$, and each power of $\omega$ beyond the leading $1/\omega$
reflects the increasing suppression of higher-order hereditary
effects in the classical limit.
 
For PN counting, each additional tail insertion brings one
more power of $GM\omega\sim v^3$ (1.5PN relative). Hence
\begin{equation}
    \text{PN order of }n\text{-tail-of-memory}
    = 2.5\text{PN} + n\times 1.5\text{PN}
    = (2.5 + 1.5n)\text{PN} \,.
\end{equation}
\begin{center}
\begin{tabular}{|l|c|c|}
\hline
\textbf{Process} & $n$ & \textbf{PN order}\\
\hline
Memory & 0 & 2.5PN\\
Tail-of-memory & 1 & 4PN\\
Tail-of-tail-of-memory & 2 & 5.5PN\\
Tail-of-tail-of-tail-of-memory & 3 & 7PN\\
$(n)$-tail-of-memory & $n$ & $(2.5+1.5n)$PN\\
\hline
\end{tabular}
\end{center}
 
In the next section we exploit the uniform IR structure $[I_1(\omega)]^n$
to resum the entire tower of amplitudes into a closed exponential form and
establish the correspondence with the Sahoo--Sen logarithmic soft
theorems~\cite{Sahoo:2018lxl,Saha:2019tub,Sen:2024qzb}.

\section{IR exponentiation, soft theorems and UV logs} \label{sec:IR_softexp}

The explicit amplitude results of the preceding sections reveal a remarkably
clean structure: the $n$-tail-of-memory amplitude is proportional to
$(GM)^n\omega^{n-1}[\mathcal{L}(\omega)]^n$ times the universal STF
memory-source tensor $\sigma^{*ij}\mathcal{Q}_{ij}$, with
$[I_1(\omega)]^n$ responsible for the entire IR and logarithmic content. In
this section we show that this structure is not accidental: summing over all
$n$ exponentiates the IR logarithm, reproduces the Sahoo--Sen soft-theorem
hierarchy, and simultaneously exposes the UV logarithms that arise from the
short-distance sensitivity of the multipole expansion.

\subsection{IR Exponentiation}

The memory amplitude ($n=0$, no $M$ insertion) has the structure:
\begin{equation}
  iA_{\rm mem} \sim \frac{i}{\omega}\,\sigma^{*ij}\mathcal{Q}_{ij}(\omega)\,,
\end{equation}
(with $\tilde{\mathcal Q}_{ij}$ defined in \eqref{eq:memory_source_tensor})
i.e.~a $1/\omega$ prefactor (the Weinberg soft pole) with no $\mathcal{L}(\omega)$
factor. The $n$-tail-of-memory amplitude ($n\geq 1$) has the structure:
\be
\ba{rcl}
  \ds iA_n & \sim&\ds \frac{(iGM)^n}{n!}\,\omega^{n-1}\,[\mathcal{L}(\omega)]^n
  \cdot\sigma^{*ij}\tilde{\mathcal{Q}}_{ij}(\omega) \\
  &=&\ds \frac{(iGM\omega\mathcal{L})^n}{n!\,\omega}
  \cdot\sigma^{*ij}\tilde{\mathcal{Q}}_{ij}(\omega).
\ea
\ee
Since the factor $(iGM\omega\mathcal{L})^n/\omega$ depends only on the
external frequency $\omega$, it multiplies the same STF source tensor at every
order. Summing over \emph{all} $n\geq 0$:
\be
\ba{rcl}
  \ds\sum_{n=0}^\infty iA_n
  &=&\ds \frac{1}{\omega}\left[\sum_{n=0}^\infty
  \frac{(iGM\omega\mathcal{L}(\omega))^n}{n!}\right]
  \cdot i\sigma^{*ij}\tilde{\mathcal{Q}}_{ij}(\omega)\\
  &=&\ds \frac{e^{iGM\omega\mathcal{L}(\omega)}}{\omega}
  \cdot i\sigma^{*ij}\tilde{\mathcal{Q}}_{ij}(\omega).
\ea
\ee
This resums the leading-IR factors. The real pole and logarithmic part of
$\mathcal{L}(\omega)$ exponentiates as a phase multiplying the STF memory
source,
while the causal $i\pi\,\mathrm{sgn}(\omega)$ term fixed by the retarded prescription gives
a real multiplicative factor and contributes to the flux. The $1/n!$ factors arise from the
attachment combinatorics, consistently with the general IR exponentiation
structure of the radiation EFT~\cite{Almeida:2021jyt}:
\begin{equation}
  iA = e^{i\phi_{IR}(\omega)/\epsilon_{IR}} \times (\text{radiative multipoles}),
  \qquad
  \phi_{IR}(\omega) = 2GE\omega\left(-\frac{\omega^2+i0^+}{\tilde\mu^2}\right)^{\epsilon_{IR}/2}.
\end{equation}
Expanding $\phi_{IR}/\epsilon_{IR}$ at $O(\epsilon^0_{IR})$ generates the $\log\omega$ term. 

A convenient choice of the IR regulator above is $\tilde \mu=1/r$, leading to $\log(\omega^2/\tilde\mu^2)\to 2\log v$. However, the actual choice of $\tilde\mu$ is irrelevant, as, e.g., shifting $\tilde \mu\to \lambda\tilde \mu$ would amount to an overall shift in the time $t\to 2GE\log\lambda$, which is unobservable \cite{Porto:2012as}.

For the \emph{pure tail} amplitude, the real pole and logarithmic part of the
exponentiated factor is a universal phase common to all multipoles and cancels
in the flux $\propto \omega^2|h(\omega)|^2$. Moreover, the absolute value of the waveform phase is not observable, but only phase differences at different times, so a universal phase drops out altogether.
On the other hand, finite terms are multipole-dependent and in
principle observable.

For memory-type
amplitudes, however, the tail phase factor multiplies $\tilde{\mathcal{Q}}_{ij}(\omega)$, which
is itself a convolution (both in frequency and time-domain) of the two quadrupoles evaluated at $\omega'$ and
$\omega-\omega'$, while the exponential depends only on their sum, which is the external frequency $\omega$.
The logarithm from the tail is \emph{local} in the frequency domain, but non-local in the time domain; hence, the tail-of-memory has a doubly non-local structure; see eq.~\eqref{eq:tom_TD}.

Diagrammatically, this is the radiation analogue of eikonal exponentiation:
repeated insertions of the background mass $M$ along the emitted-radiation line
exponentiate, rather than repeated exchanges between the hard particles~\cite{DiVecchia:2021bdo,Bjerrum-Bohr:2021vuf}.

The soft theorem hierarchy is recovered by expanding the exponential at
fixed order in $GM\omega$:
\begin{equation}
  \frac{e^{iGM\omega\mathcal{L}}}{\omega}
  = \frac{1}{\omega}\left[1 + iGM\omega\mathcal{L}
  - \frac{(GM\omega)^2}{2}\mathcal{L}^2 + \cdots\right]
\end{equation}
giving at each order the leading soft behavior:
\begin{align}
n=0:\;\frac{1}{\omega},\quad
n=1:\;GM\log(\omega r),\quad
\dots\quad
n:\;(GM)^n\omega^{n-1}(\log(\omega r))^n.
\end{align}
This hierarchy is the leading-IR pattern obtained in sections~\ref{sec:NRGR_memory} and~\ref{sec:higher}: at order $n$, the leading-IR contribution scales as $(GM)^n\sigma^{*ij}\tilde{\mathcal Q}_{ij}(0)\,\omega^{n-1}(\log\omega)^n$ in the soft limit, in correspondence with the $n$-th term of~\eqref{eq:soft_exp}~\cite{Sahoo:2018lxl,Saha:2019tub,Sen:2024qzb}. This supplies a PN/NRGR derivation of the leading-logarithmic scaling proposed in~\cite{Sen:2024qzb,Alessio:2024onn}; finite coefficients require the complete vertex contractions and PN--MPM matching. 

\subsection{Relation to Sahoo-Sen logarithmic soft theorems}

We now make explicit the correspondence between the PN/NRGR amplitude hierarchy derived above and the universal logarithmic soft-graviton theorem of Sahoo and Sen~\cite{Sahoo:2018lxl,Saha:2019tub,Sahoo:2021ctw}. The soft theorem predicts that the radiation amplitude
admits an expansion of the form
\begin{equation}
\mathcal{A}(\omega)
=
\frac{S_0}{\omega}
+
S_1 \log\omega
+
S_2 \, \omega (\log\omega)^2
+
\cdots ,
\end{equation}
where the coefficients $S_n$ are determined by the asymptotic properties of
the scattering process~\cite{Sahoo:2018lxl,Saha:2019tub,Sen:2024qzb}. Comparing with our NRGR results, the identification is direct:
 
\begin{itemize}
    \item The nonlinear (Christodoulou) memory at 2.5PN provides the hereditary contribution to $S_0/\omega$ through a one-loop amplitude with no static propagator.
    \item The tail-of-memory at 4PN provides the backscattering contribution to $S_1\log\omega$ through the single master integral $I_1(\omega)$, consistent with the Sahoo--Sen result~\cite{Sahoo:2018lxl,Saha:2019tub}.
    \item The $(n)$-tail-of-memory at $(2.5+1.5n)$PN provides the leading backscattering contribution proportional to $\omega^{n-1}(\log\omega)^n$ through $[I_1(\omega)]^n$ multiplying the finite memory form factor.
\end{itemize}

In general classical scattering, $S_1$ receives contributions from two distinct mechanisms: \emph{backscattering (tail) logarithms} (the last two terms in $B_{\mu\nu}$ of eq.~\eqref{eq:sen_tom}), from the emitted radiation scattering off the background curvature of the total mass $M$, and \emph{Coulomb-type logarithms} (the first terms in $B_{\mu\nu}$ and $C_{\mu\nu}$ of eq.~\eqref{eq:sen_tom}), from the long-time gravitational interaction among the hard particles whose asymptotic trajectories acquire corrections $x^\mu(t)\sim v^\mu t+c^\mu\log|t|$. A natural framework for making this decomposition precise is provided by \emph{logarithmic translations}~\cite{Boschetti:2025tru} — residual gauge freedoms $\xi^\mu\sim\log R\,L^\mu$ that shift the log deviation vector of asymptotic geodesics by a constant while leaving the full log soft factor invariant. This gauge freedom determines how the backscattering and Coulomb contributions are distributed between the drag phase and the hard particle angular momenta depending on the choice of log frame, but the total gauge-invariant soft factor remains the same~\cite{Boschetti:2025tru}.
While in general $S_1 = S_1^{\rm Coulomb} + S_1^{\rm tail}$~\cite{Laddha:2018vbn}, in the PN bound-system framework, the hard particles never reach asymptotic free states — they remain gravitationally bound for all time — so their logarithmic trajectory deviations $c^\mu_i$ are not defined and the Coulomb contribution to $S_1$ is absent. Only the backscattering (drag) contribution survives, since the emitted radiation does propagate to null infinity and its log deviation is well-defined. The same holds at all orders: the leading logarithms $\omega^{n-1}(\log\omega)^n$ are generated solely by the $n$-tail ladder.

\subsection{Subtraction procedure and isolation of UV divergences}

While the IR logarithms $\log^n(\omega)$ are controlled entirely by the $\sigma\sigma\phi$ polarization sector, the full amplitude at each order $n\geq 2$ also receives contributions from sectors involving the gravito-magnetic field $A_i$
and scalar $\phi$. These do not generate soft logarithms (see Appendix~\ref{sec:app_polarsector}), but introduce UV divergences that must be renormalized. At $n=1$ the $\{A\sigma\}$ sector contributes only finite constants; at $n\geq 2$ it produces UV poles $\alpha^{(n)}/\epsilon_{UV}$ generating UV logarithms.
Physically, the UV divergences reflect the breakdown of the point-like multipolar description below the short-distance orbital scale $r$. 

After regularization and renormalization, they correspond to logarithms $\log(\omega r)$ encoding the matching to the near-zone dynamics \cite{Foffa:2019yfl}.
At each order $n\geq 2$, the single IR pole predicted by exponentiation of lower-order results can be subtracted, isolating the UV pole~\cite{Almeida:2021jyt}.

The isolated UV pole is then absorbed into the bare multipole operator\footnote{In \eqref{eq:bare_renormalized} the bare multipole moment does not depend on the renormalization scale $\mu$,
but $G$ does, according to $\mu \, dG/d\mu=-\epsilon G$.}:
\begin{equation}
  I^{iji_1\cdots i_r}_B(\omega)
  = \left[1 - \frac{\beta(r)}{2\epsilon_{UV}}(GM\omega)^2\right]
  I^{iji_1\cdots i_r}_R(\omega,\mu)\,,
  \label{eq:bare_renormalized}
\end{equation}
where $I_B^{iji_1\cdots i_r}$ is the bare source multipole, $I_R^{iji_1\cdots i_r}(\omega,\mu)$ is the corresponding renormalized multipole at the scale $\mu$,
and $\beta(r)$ is the $2^{r+2}$-multipole-dependent beta function. It determines the associated renormalization-group running and hence the UV logarithms~\cite{Almeida:2021jyt,Ivanov:2025ozg}. These logarithms come from renormalization of a single multipole operator, starting at $O(G^2)$
~\cite{Almeida:2021jyt,Ivanov:2025ozg}, and hence subleading with respect to the IR logs.
\section{Conclusions}
\label{sec:conclusions}
 
In this paper we have studied the soft structure of gravitational waveforms generated
by compact binaries within the multipolar post-Newtonian framework, using the NRGR
effective field theory to compute the relevant Feynman amplitudes explicitly. Our
central finding is an explicit correspondence between the hierarchy of
hereditary radiation effects — memory, tail-of-memory, and their higher-order
generalisations — and the universal logarithmic soft-graviton theorem of Sahoo, Sen
and collaborators~\cite{Sahoo:2018lxl,Saha:2019tub,Sahoo:2021ctw}, as reviewed
in~\cite{Sen:2024qzb}.
 
\paragraph{Memory and tail-of-memory.}
The nonlinear (Christodoulou) memory at 2.5\,PN arises from a one-loop
amplitude with two retarded propagators and no static propagator. Its TT radiative
component has no IR divergence or hereditary logarithm and carries only the
Weinberg $1/\omega$ pole~\cite{Weinberg:1965nx} — the PN realisation of the leading
displacement-memory coefficient $\tilde{h}_{\mu\nu}^{(0)}$ of the universal soft
expansion~\eqref{eq:soft_exp}~\cite{He:2014laa,Strominger:2014pwa,Sen:2024qzb}. The
tail-of-memory at 4\,PN involves the addition of a single static propagator, generating one IR (unphysical) pole
$1/\epsilon_{\rm IR}$ and the associated (physical) $G M\omega\log\omega$ factor. In the soft 
limit, the amplitude behaves as $\sim G M \cdot E_{\rm rad} \cdot \ln\omega$, in
precise agreement with the subleading Sahoo--Sen coefficient
$\tilde{h}_{\mu\nu}^{(\log)}$~\cite{Sahoo:2018lxl,Saha:2019tub,Laddha:2018vbn}.

\paragraph{Spin memory.}
For completeness, we have also incorporated the magnetic-parity spin-memory
observable.  In contrast to displacement memory, it is encoded in the change of the retarded-time-integrated magnetic-parity metric perturbation (momentum-octupole) and is measured as the relative time delay of counter-orbiting null
rays~\cite{Pasterski:2015tva,Nichols:2017rqr}.  Its classical soft origin is the ordinary
subleading soft graviton theorem: the energy flux entering displacement memory is
replaced by angular-momentum flux. The resulting spin-memory contribution is a finite, analytic $\omega^{0}$ term.  Unlike the nonanalytic $1/\omega$
displacement-memory term, it can therefore receive contributions from local
dynamics at the same soft order.  Consequently, the waveform obtained from the
angular-momentum-flux soft kernel does not by itself coincide with the instantaneous PN-MPM current-octupole waveform~\cite{Blanchet:1997ji}.  The complete $\ell=3$ spin-memory can be obtained from the soft flux formula by including the superrotation contribution.
 
\paragraph{All-order leading-logarithm tower.}
The computation generalises to the process with $n$ static mass insertions, the
$(n)$-tail-of-memory at PN order $(2.5 + 1.5n)$. Each additional mass insertion
contributes one further factor of the master integral $I_1(\omega)$, adding one power
of $\mathcal L(\omega) = 1/\epsilon_{\rm IR} - \tfrac{1}{2}\ln(\omega^2/\tilde\mu^2) + \cdots$
to the amplitude. The resulting $n$-fold IR pole and physical $\ln^n\omega$ yield the
universal soft behaviour
\begin{equation}
  \mathcal{A}_n\big|_{\omega\to 0}
  \;\sim\; (G M)^n \cdot E_{\rm rad} \cdot \omega^{n-1}(\ln\omega)^n.
  \label{eq:ntail_soft}
\end{equation}
This reproduces the $n$-th level of the Sahoo--Sen
hierarchy~\cite{Sahoo:2018lxl,Saha:2019tub,Sahoo:2021ctw} at leading IR order. The PN loop-counting is transparent: each
$\sigma\sigma\phi$ vertex brings $\omega^2$, each static propagator integral
$I_1(\omega)$ returns $\log(\omega)/\omega$; the remaining finite coefficient is fixed
by the complete vertex contraction and PN--MPM matching, giving the net
$\omega^{n-1}$ factor.
 
\paragraph{IR exponentiation and soft dressing.}
Resumming the tower of $n$-tail-of-memory amplitudes, we find
\begin{equation}
  \sum_{n=0}^\infty G^n\mathcal{A}_n
  \;=\; \frac{e^{i G M \omega {\mathcal L}(\omega)}}{\omega}
    \cdot G\int\!\frac{d\omega'}{2\pi}\,\omega'^2
    i\mathcal{A}_0(\omega-\omega')\cdot i\mathcal{A}_0(\omega'),
\end{equation}
where $\mathcal{A}_0$ is the leading quadrupole amplitude and the convolution is
proportional to the total radiated energy $E_{\rm rad}$. This is consistent with the
general IR exponentiation structure of the radiation EFT~\cite{Almeida:2021jyt} and
the resummation of universal tails. For memory-type amplitudes, which are bilinear in
$\mathcal{A}_0$, the logarithms from the tails are \emph{local} in the frequency domain, but non-local in the time domain; hence in the tail-of-memory we have a doubly non-local structure. 
 
\paragraph{Distinction between IR and UV logarithms.}
UV divergences first appear at $n=2$ and originate from the $\{A\sigma\}$
polarisation sector. After renormalisation via the subtraction procedure
of~\cite{Almeida:2021jyt}, these poles generate UV logarithms $\ln(\omega r)$, corresponding to a physical scale dependence of the multipole in the far theory.
The IR logarithms appear in the waveform phase, inducing a multipole-independent logarithmic term in the waveform phase.

\paragraph{Coulomb versus backscattering logarithms.}
In a general classical scattering, the subleading soft coefficient receives contributions
from two sources~\cite{Sahoo:2018lxl,Saha:2019tub,Laddha:2018vbn,Sen:2024qzb}:
Coulomb-type logarithms from the long-time gravitational interaction among the hard
particles, and backscattering logarithms from radiation propagating through the static
background curvature (the tail sector). In the PN bound-system framework, the hard particles never reach asymptotic free states and their logarithmic trajectory deviations are not defined, so the Coulomb contribution to $S_1$ is absent — the explicit IR logarithms arise solely from the backscattering (tail) sector. This contrasts with the hyperbolic PM case where both Coulomb and backscattering contributions are present~\cite{Laddha:2018vbn,Alessio:2024onn}.
 
\paragraph{Physical implications and remarks.}
An important observation from the classical soft-theorem perspective~\cite{Sen:2024qzb}
is that for a \emph{binary black hole merger} — where the final state contains a single
black hole — the tail coefficients $B^{\mu\nu}$ and $F^{\mu\nu}$ vanish identically,
so no late-time logarithmic tails are present. In contrast, for bound inspiralling
binaries modelled by the PN--MPM framework, the hereditary structure is characterised
by the multipole moments, and the logarithmic IR structure computed in this paper
enters directly through the radiation EFT. These logarithmic tail corrections to the
waveform are genuine observables affecting the waveform phase at $O(G)$ \cite{Blanchet:1994ez}, but they drop out from the flux (unlike UV logs which start at $O(G^2)$).
 
\paragraph{Outlook.}
The present work opens several directions for future investigation. On the side of extending the NRGR computation, it would be natural to go beyond the electric quadrupole sector to the magnetic quadrupole, electric octupole, and spin-dependent contributions — in particular the spin-quadrupole tail at 4\,PN \cite{Trestini:2023wwg} and its associated tail-of-memory at 6.5\,PN~\cite{Ghosh:2021bam}.  An important next step is a direct NRGR calculation of spin memory from the current-multipole sector.  This would realise the subleading infrared triangle between angular-momentum flux, the Cachazo--Strominger soft theorem, and BMS superrotations~\cite{Campiglia:2014yka,Campiglia:2015yka,Pasterski:2015tva}, and would make the relation to the retarded-time delay observable quantitative.
 
On the side of connecting to PM and amplitude methods, explicit comparison with the logarithmic PM waveforms of~\cite{Alessio:2024onn,Bautista:2021wfy} at small velocities would sharpen the Coulomb/tail decomposition of $S_1$, and making the KMOC-based analysis of~\cite{Bautista:2021llr,Akhtar:2024lkk,Aoki:2026fgt} directly comparable to the NRGR computation performed here would be valuable.
The present results also provide a concrete PN realisation of one edge of the infrared triangle connecting the logarithmic soft theorem~\cite{Sahoo:2018lxl,Saha:2019tub}, 
and the memory and its tails interpreted as sub-leading soft corrections.

Making this explicit at the level of asymptotic charges of the BMS group would complete a second side of the infrared triangle,  relating the memory effect to the symmetries at null infinity, represented by the BMS group.
At null infinity, a \emph{supertranslation} shifts the non-oscillatory part of the strain. Thus, two waveforms related by a relative supertranslation do not have the same strain baseline at equal coordinate values of $u$ and $(\theta,\phi)$.  
Correct identification of the BMS frame is necessary for comparing waveforms obtained within different approximation schemes, e.g., when matching PN and numerical-relativity waveforms~\cite{Mitman:2024uss}.  The distinct non-oscillatory strain baselines label degenerate non-radiative BMS vacua.  The passage of radiation takes the spacetime from the early to the late vacuum: such a vacuum transition is interpreted as the displacement memory, and the Ward identity of the associated supertranslation symmetry yields Weinberg's leading soft graviton theorem, representing the last edge of the \emph{infrared} triangle, which overall relates asymptotic symmetries, memory effects, and soft theorems.

A particularly important open problem is the formulation of an S-matrix framework for gravitationally
\emph{bound} systems that would allow quantum soft theorems — corresponding to the
tail-of-memory and higher-order tail insertions — to be derived in close analogy with the hyperbolic scattering case. In
the hyperbolic case the standard S-matrix applies: the initial and
final states are well-defined asymptotic states at $t\to\pm\infty$, and quantum soft graviton theorems follow as Ward identities of
BMS symmetries~\cite{He:2014laa,Strominger:2014pwa,Strominger:2016wns}. For bound systems this faces a fundamental obstruction: the binary never reaches a free asymptotic state, so conventional in/out Fock-space scattering states do not apply. The initial and final states must be defined differently —
through adiabatic switching, in-in (Schwinger--Keldysh) expectation
values~\cite{Almeida:2021jyt}, or dressed asymptotic states encoding the permanent hereditary effects. Developing such a framework would put the classical results here on a firmer quantum-field-theoretic footing and clarify whether BMS Ward identities persist for bound sources. 
 
\acknowledgments
SA thanks Alok Laddha and Akavoor Manu for helpful discussions and feedback on the draft. SA is supported by the São Paulo Research Foundation (FAPESP) under Grant No. 2025/01291-6.
RS is supported by FAPESP Grant n. 2022/03650 and 2021/14335-0, as well as by CNPq grant 309659/2025-6.

\begin{appendix}

\section{Frequency and angular integrals in memory kernels}
\label{sec:app_memory_angular}

This appendix collects the frequency-region analysis and the sphere
integrals used in the memory, tail-of-memory, and spin-memory derivations.
We first isolate the nonanalytic part of the tail-of-memory convolution and
then derive the angular kernels that project the relevant multipolar
components of the energy and angular-momentum fluxes.

\subsection{Hard and soft regions in the tail-of-memory convolution}
\label{sec:app_tom_regions}

Starting from the symmetrized frequency convolution
\eqref{eq:tom_symmetrized}, we denote its logarithmic bracket by
\begin{equation}
\mathcal B(\omega,\omega')
\equiv \omega^\prime\ln|\omega-\omega^\prime|
+(\omega-\omega^\prime)\ln|\omega^\prime|.
\label{eq:app_tom_bracket}
\end{equation}
The hard and soft regions isolate, respectively, the analytic and
nonanalytic dependence on the external frequency.

In the hard region $ |\omega^\prime|\gg |\omega| $, one may write
\begin{align}
\ln|\omega-\omega^\prime|
&=\ln|\omega^\prime|+\ln\left|1-\frac{\omega}{\omega^\prime}\right|
\nonumber\\
&=\ln|\omega^\prime|-\frac{\omega}{\omega^\prime}
-\frac{\omega^2}{2(\omega^\prime)^2}
+\mathcal O\!\left(\frac{\omega^3}{(\omega^\prime)^3}\right).
\label{eq:app_tom_hard_log}
\end{align}
Substitution into \eqref{eq:app_tom_bracket} gives
\begin{equation}
\mathcal B(\omega,\omega')
=\omega\bigl[\ln|\omega^\prime|-1\bigr]
-\frac{\omega^2}{2\omega^\prime}
+\mathcal O\!\left(\frac{\omega^3}{(\omega^\prime)^2}\right).
\label{eq:app_tom_hard}
\end{equation}
The memory kernel
$\mathcal K_{ij}(\omega^\prime,\omega-\omega^\prime)$ has an ordinary
Taylor expansion in $\omega$ about
$(\omega^\prime,-\omega^\prime)$.  The hard-region integrand therefore
has a power expansion in the external frequency. In particular, it produces
no $\ln|\omega|$.

In the soft region, we set $\omega^\prime=\omega x$.  For the
positive-frequency radiative contribution $0<x<1$, $d\omega^\prime
=\omega\,dx$, and \eqref{eq:app_tom_bracket} becomes
\begin{equation}
\mathcal B(\omega,\omega x)
=\omega\ln|\omega|+\omega\,\phi(x),\qquad
\phi(x)\equiv x\ln(1-x)+(1-x)\ln x.
\label{eq:app_tom_soft_kernel}
\end{equation}
Consequently, the soft part of \eqref{eq:tom_symmetrized} is
\begin{align}
\left.\tilde{\mathcal U}_{ij,4\,\mathrm{PN}}^{\mathrm{t.o.m.}}(\omega)
\right|_{\rm soft}
&\propto\frac12\ln|\omega|
\int_0^1\frac{dx}{2\pi}\,
\mathcal K_{ij}(\omega x,\omega(1-x))\nonumber\\
&\quad+\frac12\int_0^1\frac{dx}{2\pi}\,
\mathcal K_{ij}(\omega x,\omega(1-x))\,\phi(x).
\label{eq:app_tom_soft_decomposition}
\end{align}
The second term is analytic in $\omega$.  Indeed, the memory kernel is
analytic, and every coefficient in its expansion gives integrals of the
form
\begin{equation}
\int_0^1 dx\,x^m(1-x)^n\phi(x),\qquad m,n\geq0.
\label{eq:app_tom_endpoint_integrals}
\end{equation}
They are finite because
$\phi(x)=\ln x$ as $x \to 0$, while
$\phi(x)=\ln(1-x)$ as $x \to 1$.
Thus the first term in \eqref{eq:app_tom_soft_decomposition} is the sole
nonanalytic contribution.  Equivalently, the nonanalytic
$\omega\ln|\omega|$ term in the bracket \eqref{eq:app_tom_soft_kernel}
is the tail correction to the memory kernel; the prefactor
$1/\omega$ in \eqref{eq:tom_symmetrized} yields the resulting
$\ln|\omega|$ behavior of the tail-of-memory.

\subsection{Angular integral in the memory kernel}

\paragraph{$\ell=2$ component.}
This subsection derives the angular identity used in
\eqref{eq:memory_kernel_l2},
\begin{equation}
\mathcal K_{ij,mn}(\hat n)
\equiv \int d\Omega'\,
\frac{\Pi_{ijkl}(\hat n)\hat n_k'\hat n_l'}{1-\hat n\cdot\hat n'}\,
\hat n'_{\langle mn\rangle}
=\frac{2\pi}{3}\Pi_{ijmn}(\hat n)\,,
\label{eq:app_memory_kernel_l2}
\end{equation}
with projectors defined in \eqref{eq:proj_def}.

The integral is invariant under rotations about $\hat n$ and is parity
even.  Its action on the transverse STF $\ell=2$ subspace is therefore
proportional to the identity on that subspace;
$\mathcal K_{ij,mn}=C\Pi_{ijmn}$, for a scalar coefficient $C$.

To determine $C$, we take $\hat n=\hat z$, for which
$P_{ij}(\hat z)=\mathrm{diag}(1,1,0)$, and conveniently
use the transverse STF tensor
\begin{equation}
E_{ij}\equiv \mathrm{diag}(1,-1,0)\,,
\end{equation}
which obeys $P_{ik}(\hat z)E_{kj}=E_{ij}$, $P_{ij}(\hat z)E_{ij}=0$, and
$E_{ij}E_{ij}=2$, and hence
$E_{ij}\Pi_{ijmn}(\hat z)E_{mn}=E_{ij}E_{ij}=2$.  Parameterizing the integration
direction as
\begin{equation}
\hat n_i'=\big(\sqrt{1-x^2}\cos\phi,\sqrt{1-x^2}\sin\phi,x\big),
\qquad x=\cos\theta,\qquad d\Omega'=dx\,d\phi,
\end{equation}
one has
\begin{align}
E_{ij}\Pi_{ijkl}(\hat z)\hat n_k'\hat n_l'
&= E_{kl}\hat n_k'\hat n_l'
 =\hat n_x'^2-\hat n_y'^2
 =(1-x^2)\cos(2\phi), \nonumber\\
E_{mn}\hat n'_{\langle mn\rangle}
&=E_{mn}\hat n_m'\hat n_n'
 =\hat n_x'^2-\hat n_y'^2
 =(1-x^2)\cos(2\phi),
\end{align}
where the trace subtraction in $n'_{\langle mn\rangle}$ drops out because
$E_{mm}=0$.  Contracting 
\eqref{eq:app_memory_kernel_l2} with $E_{ij}E_{mn}$ then gives 
\begin{align}
E_{ij}\mathcal K_{ij,mn}E_{mn}
&=\int d\Omega'\,
\frac{(1-x^2)^2\cos^2(2\phi)}{1-x}
\nonumber\\
&=\left[\int_{-1}^{1}dx\,(1+x)(1-x^2)\right]
\left[\int_0^{2\pi}d\phi\,\cos^2(2\phi)\right] \nonumber\\
&=\frac{4\pi}{3},
\label{eq:app_memory_kernel_contraction}
\end{align}
where
\begin{equation}
\int_{-1}^{1}dx\,(1+x)(1-x^2)=\frac43,\qquad
\int_0^{2\pi}d\phi\,\cos^2(2\phi)=\pi.
\end{equation}
Now short-circuiting
$\mathcal K_{ij,mn}=C\Pi_{ijmn}$ and 
$E_{ij}\mathcal K_{ij,mn}E_{mn}=2C$, equation
\eqref{eq:app_memory_kernel_contraction} fixes
$C=2\pi/3$, proving \eqref{eq:app_memory_kernel_l2}. 
\paragraph{$\ell=0$ and $\ell=4$ components.}
The $\ell=0$ component provides the luminosity.  Using the standard
isotropic angular averages for STF tensors~\cite{Thorne:1980ru},
\begin{equation}
\int d\Omega'\,n_i'n_j'=\frac{4\pi}{3}\delta_{ij},\qquad
\int d\Omega'\,n_i'n_j'n_k'n_l'
=\frac{4\pi}{15}
\bigl(\delta_{ij}\delta_{kl}+\delta_{ik}\delta_{jl}
+\delta_{il}\delta_{jk}\bigr),
\label{eq:app_isotropic_averages}
\end{equation}
in the scalar contraction in \eqref{eq:FL} gives
\begin{align}
\int d\Omega'\,
\Pi_{ijkl}(\hat n')\mathcal M_{ij}^{(3)}\mathcal M_{kl}^{(3)} \Big|_{l=0}
={}&4\pi\mathcal M_{ab}^{(3)}\mathcal M_{ab}^{(3)}
-\frac{8\pi}{3}\mathcal M_{ab}^{(3)}\mathcal M_{ab}^{(3)}
+\frac{4\pi}{15}\mathcal M_{ab}^{(3)}
\mathcal M_{ab}^{(3)}\nonumber\\
={}&\frac{8\pi}{5}\mathcal M_{ab}^{(3)}
\mathcal M_{ab}^{(3)}.
\label{eq:app_monopole_luminosity_integral}
\end{align}

Hence
\begin{equation}
\left.\frac{dE^Q}{du}\right|_{\ell=0}
=\frac{G}{5c^5}\mathcal M_{ab}^{(3)}
\mathcal M_{ab}^{(3)}.
\label{eq:app_monopole_luminosity}
\end{equation}
The TT memory kernel removes this isotropic component:
\begin{equation}
\int d\Omega'\,
\frac{\Pi_{ijkl}(\hat n)n_k'n_l'}{1-\hat n\cdot\hat n'}=0.
\label{eq:app_memory_kernel_l0}
\end{equation}

The $\ell=4$ part of the luminosity is
\begin{equation}
\left.\frac{dE^Q}{du\,d\Omega'}\right|_{\ell=4}
=\frac{G}{16\pi c^5}\,
\mathcal M_{\langle ab}^{(3)}\mathcal M_{cd\rangle}^{(3)}
n'_{\langle abcd\rangle}\,,
\label{eq:app_quadrupole_flux_l4}
\end{equation}
and its corresponding angular integral in the TT memory kernel is
\begin{equation}
\int d\Omega'\,
\frac{\Pi_{ijkl}(\hat n)\hat n_k'\hat n_l'}{1-\hat n\cdot\hat n'}\,
n'_{\langle abcd\rangle}
=\frac{4\pi}{15}\Pi_{ij\langle ab}(\hat n)\hat n_c\hat n_{d\rangle}.
\label{eq:app_memory_kernel_l4}
\end{equation}

Substitution of
\eqref{eq:app_quadrupole_flux_l4} into the memory formula consequently gives
\begin{equation}
\left.h_{ij}^{\rm mem,Q}(u,\hat n)\right|_{\ell=4}
=\frac{G^2}{15Rc^9}\int_{-\infty}^{u}du'\,
\mathcal M_{\langle ab}^{(3)}(u')
\mathcal M_{cd\rangle}^{(3)}(u')\,
\Pi_{ij\langle ab}(\hat n)\hat n_c\hat n_{d\rangle}.
\label{eq:app_memory_hexadecapole}
\end{equation}
Thus the $\ell=4$ component produces the radiative mass
hexadecapole, while the $\ell=2$ calculation above is the projection
relevant for the radiative mass quadrupole.

\subsection{Angular integrals in the spin-memory kernel}
\label{sec:app_spin_memory_angular}

We first verify the normalization of the ordinary $\ell=1$
angular-momentum luminosity.  With
$E_i\equiv\epsilon_{iab}\mathcal M_{ac}^{(2)}
\mathcal M_{bc}^{(3)}$, the two angular integrals in
\eqref{eq:ang_momentum_flux_density} reduce to
\begin{align}
&\int d\Omega'\,\epsilon_{ikl}
\mathcal M_{cd}^{(2)}\mathcal M_{fg}^{(3)}\hat n'_k
\left(\hat n'_f\Pi_{cdlg}(\hat n')
+\hat n'_g\Pi_{cdlf}(\hat n')\right)
=\frac{16\pi}{15}E_i,
\label{eq:app_spin_orbital_integral}\\
&\int d\Omega'\,\epsilon_{ikl}
\mathcal M_{mn}^{(2)}\mathcal M_{cd}^{(3)}
\Pi_{almn}(\hat n')\Pi_{akcd}(\hat n')
=\frac{16\pi}{15}E_i\,.
\label{eq:app_spin_helicity_integral}
\end{align}
For the first
line, one expands a single TT projector and uses
\begin{equation}
\int d\Omega'\,\Pi_{ijkl}(\hat n')
=\frac{2\pi}{15}
\left(11\delta_{ik}\delta_{jl}-4\delta_{ij}\delta_{kl}
+\delta_{il}\delta_{jk}\right).
\label{eq:app_spin_single_projector_average}
\end{equation}
For the second, the product is first reduced according to
\begin{equation}
\Pi_{almn}\Pi_{akcd}
=\frac12\left(P_{ln}\Pi_{mkcd}-P_{mn}\Pi_{lkcd}\right),
\qquad P_{ij}=\delta_{ij}-\hat n'_i\hat n'_j,
\label{eq:app_spin_projector_product}
\end{equation}
after which the same rank-two and rank-four sphere averages give
\eqref{eq:app_spin_helicity_integral}.  Including the prefactors in
\eqref{eq:ang_momentum_flux_density} yields
\begin{equation}
\frac{dJ_i}{du}= \left(\frac{2}{15}  + \frac{4}{15}\right)\frac{G}{c^5}E_i  = \frac{2G}{5c^5}E_i,
\label{eq:app_spin_orbital_helicity_result}
\end{equation}
which proves \eqref{eq:quadrupolar_ang_momentum_luminosity}. 

\paragraph{$\ell=1$ component.}

The angular integral associated with the $\ell=1$ angular-momentum flux is
\begin{equation}
\mathcal K^{(0)}_{ijk}(\hat n)\equiv
\Pi_{ijkl}(\hat n)\int d\Omega'\,
\frac{\hat n'^l}{1-\hat n\!\cdot\hat n'}.
\end{equation}
The unprojected integral has a longitudinal forward singularity.  Since
$\Pi_{ijkl}(\hat n)\hat n_l=0$, subtracting its longitudinal part gives
\begin{align}
\mathcal K^{(0)}_{ij,k}
&=\Pi_{ijkl}(\hat n)\int d\Omega'\,
\frac{\hat n'^l-\hat n^l}{1-\hat n\!\cdot\hat n'},\nonumber\\
\int d\Omega'\,\frac{\hat n'^l-\hat n^l}{1-\hat n\!\cdot\hat n'}
&=-4\pi \hat n^l.
\end{align}
Thus $\mathcal K^{(0)}_{ij,k}=0$.

\paragraph{$\ell=3$ component.}

We define
\begin{equation}
\mathcal K_{ijkab}(\hat n)\equiv
\Pi_{ijkl}(\hat n)\int d\Omega'\,
\frac{\hat n'^l \hat n'_{\langle ab\rangle}}
{1-\hat n\!\cdot\hat n'}.
\end{equation}
Put $x=\hat n\!\cdot\hat n'$ and write
\begin{equation}
\hat n'^i=x\hat n^i+\sqrt{1-x^2}\,e^i(\phi),\qquad
\hat e^i\hat n_i=0,\qquad \hat e^i\hat e_i=1,\qquad d\Omega'=dx\,d\phi,
\end{equation}
where $\hat e^i(\phi)$ is the unit azimuthal vector in the plane transverse to
$\hat n$.  In a frame with $\hat n=\hat z$, it is
$\hat e^i(\phi)=(\cos\phi,\sin\phi,0)$.
The TT projector removes the longitudinal part of $\hat n'^l$.  Using
\begin{equation}
\int_0^{2\pi}\!d\phi\,\hat e^l\hat e^a=\pi P^{la}(\hat n),\qquad
\int_0^{2\pi}\!d\phi\,\hat e^l\hat e^a\hat e^b=0,
\end{equation}
the only surviving term is proportional to
$x(1-x^2)\hat e^l(\hat n_a\hat e_b+\hat n_b\hat e_a)$.  Therefore
\begin{align}
\mathcal K_{ijkab}
&=\pi\Pi_{ijkl}(\hat n)
\left(\hat n_aP_{lb}(\hat n)+\hat n_bP_{la}(\hat n)\right)
\int_{-1}^{1}\!dx\,\frac{x(1-x^2)}{1-x}\nonumber\\
&=\frac{2\pi}{3}
\left(\Pi_{ijkb}(\hat n)\hat n_a+\Pi_{ijka}(\hat n)\hat n_b\right)\nonumber\\
&=\frac{4\pi}{3}\Pi_{ijk\langle a}\hat n_{b\rangle},
\end{align}
where $\int_{-1}^{1}dx\,x(1+x)=2/3$.  This proves
\eqref{eq:spin_memory_kernel_L2}. 

\paragraph{$\ell=5$ component.}

For completeness, we consider
\be
\mathcal K^{(4)}_{ijkabcd}(\hat n)
\equiv\Pi_{ijkl}(\hat n)\int d\Omega'\,
\frac{\hat n'^l \hat n'_{\langle abcd\rangle}}{1-\hat n\cdot\hat n'}.
\ee
Contract it with an arbitrary rank-four STF tensor $X_{abcd}$.  Since
$X_{abcd}\delta_{ab}=0$,
\be
X_{abcd}\hat n'_{\langle abcd\rangle}
=X_{abcd}\hat n'_a\hat n'_b\hat n'_c\hat n'_d.
\ee
With $x=\hat n\cdot\hat n'$ and
\be
\hat n'^i=x\hat n^i+\sqrt{1-x^2}\,\hat e^i(\phi),
\qquad \hat e^i\hat n_i=0,
\ee
the TT projector selects the transverse part of $n'^l$.  The azimuthal
average is nonzero only for one or three factors of $e^i$ among the four
directions contracted with $X_{abcd}$.  Define
\be
\mathcal X_{ij,k}\equiv
\Pi_{ijka}X_{abcd}\hat n_b\hat n_c\hat n_d.
\ee
The term with one transverse vector gives
\be
\left.X_{abcd}\mathcal K^{(4)}_{ijkabcd}\right|_{e^1}
=4\pi\mathcal X_{ij,k}\int_{-1}^{1}dx\,
\frac{x^3(1-x^2)}{1-x}
=\frac{8\pi}{5}\mathcal X_{ij,k}.
\ee
For the term with three transverse vectors, we use
\be
\int_0^{2\pi}\!d\phi\,\hat e^l\hat e^a\hat e^b\hat e^c
=\frac{\pi}{4}\left(P^{la}P^{bc}+P^{lb}P^{ac}+P^{lc}P^{ab}\right).
\ee
As $X_{abcd}P_{bc}=-X_{abcd}\hat n_b\hat n_c$, this gives
\be
\begin{aligned}
\left.X_{abcd}\mathcal K^{(4)}_{ijkabcd}\right|_{e^3}
&=3\pi\mathcal X_{ij,k}\int_{-1}^{1}dx\,
\frac{x(1-x^2)^2}{1-x}\\
&=-\frac{4\pi}{5}\mathcal X_{ij,k}.
\end{aligned}
\ee
The two contributions therefore combine to
\be
X_{abcd}\mathcal K^{(4)}_{ijkabcd}
=\frac{4\pi}{5}\Pi_{ijka}X_{abcd}\hat n_b\hat n_c\hat n_d.
\ee
Since $X_{abcd}$ is arbitrary and STF, this proves
\be
\mathcal K^{(4)}_{ijkabcd}
=\frac{4\pi}{5}\Pi_{ijk\langle a}\hat n_b\hat n_c\hat n_{d\rangle},
\ee
which is \eqref{eq:spin_memory_kernel_L4}.

\section{The NRGR Framework} \label{sec:app_nrgr}

In this appendix we collect the elements of the NRGR (Non-Relativistic General Relativity) effective field theory of Goldberger and Rothstein~\cite{Goldberger:2004jt} that are used in the main text. The presentation follows the conventions of~\cite{Almeida:2021jyt,Foffa:2019eeb}. The framework organises the metric in the Kaluza-Klein (KK) decomposition and generates the Feynman rules for graviton emission from a multipolar point-particle source; we specialise throughout to the radiation zone where $|\mathbf{k}|\sim\omega$.

\subsection{Kaluza-Klein Decomposition}

The NRGR framework begins with the Kaluza-Klein (KK) decomposition of
the metric. Instead of working with the full metric $g_{\mu\nu}$, we
decompose it into scalar, vector, and tensor modes:
\begin{equation}
    g_{\mu\nu} \to \{\phi, A_i, \sigma_{ij}\}
\end{equation}
where $\phi$ is the scalar (Newtonian potential), $A_i$ is the
gravito-magnetic vector, and $\sigma_{ij}$ is the tensor (graviton) mode.
This decomposition is adapted to the non-relativistic nature of the source.

The metric is written as:
\begin{equation}
    ds^2 = -e^{2\phi/\Lambda}\pa{dt + \frac{A_i}\Lambda dx^i}^2 +
    e^{-2\phi/((d-2)\Lambda)}\gamma_{ij}dx^idx^j
\end{equation}
where $\gamma_{ij} = \delta_{ij}+\sigma_{ij}/\Lambda$ encodes the
tensor perturbation and
\begin{equation}
    \Lambda\equiv \paq{32\pi G}^{-1/2}=\paq{32\pi G_N\tilde\mu^{3-d}}^{-1/2}\,,
\end{equation}
has been introduced to endow the gravitational fields with canonical mass dimension. Note that for dimensional regularization we need the Newton's constant $G$ in generic $d+1$ dimensions, related to the $3+1$ Newton's constant $G_N$ by the arbitrary inverse length scale $\tilde \mu$.
 In the weak-field limit, expanding to the relevant
order in the power of the KK fields gives interactions that appear in the Feynman rules.

\subsection{The Bulk Action}

The gravitational bulk action is the Einstein-Hilbert action:
\begin{equation}
    S_{\rm bulk} = \frac{1}{16\pi G}\int d^{d+1}x\,\sqrt{-g}\,R\,,
\end{equation}
plus a gauge-fixing term.
For the tail-of-memory calculation, the two source multipoles meet at the
complete $\sigma\sigma\sigma$ vertex $V_1$.  Its source-contracted tensor
numerator is retained without scalar reduction and is denoted by
$\mathcal V_{1\,ab\,cd\,kl}(\omega_1,\mathbf q;\omega_2,\mathbf p-\mathbf q)$
in \eqref{eq:tom_memory_subgraph}.  This is essential: only after the
$\mathbf q$ integration and the TT projection does the resulting
$\mathcal H^{\rm mem}_{kl}(\omega,\mathbf p)$ have an on-shell limit.

The second cubic vertex couples this intermediate $\sigma$ mode to the
static $\phi$ field sourced by the mass insertion.  Its projected
contraction is
\begin{equation}
 P_{\sigma_{kl}\sigma_{ab}}V^{(3)}_{2\,abij}
 =-2i\,\frac{\omega^2}{\Lambda}\,g^{i(k}g^{l)j},
 \label{eq:app_PssV2}
\end{equation}
in agreement with \eqref{eq:PssV2}.  Thus the outer tail loop has a
$\mathbf p$-independent $\omega^2$ numerator at the on-shell pinch singularity.  The remaining
polarization sectors have momentum factors that remove this IR singularity;
their role in the IR/UV separation is summarized in
Appendix~\ref{sec:app_polarsector}.  The complete KK rules used here follow
from~\cite{Foffa:2013qca,Almeida:2024lbv}.

\subsection{The Multipole Action}

The source action describes how the compact binary couples to the
gravitational field via its multipole moments. The worldline action is:
\begin{equation}
    S_{\rm p} = \int dt\left[-M + \frac{1}{2}{\mathcal I}^{ij}\mathcal{E}_{ij}
    + \frac{1}{2}{\mathcal J}^{ij}\mathcal{B}_{ij} + \cdots\right]
\end{equation}
where $\mathcal{E}_{ij}$ and $\mathcal{B}_{ij}$ are the electric and
magnetic parts of the Weyl tensor evaluated on the worldline.

For generic electric $2^{r+2}$-pole moments the coupling is:
\begin{equation}
    S_{\rm p} \supset \frac{1}{2}\int dt\, c_r^{(I)}
    {\mathcal I}^{iji_1\cdots i_r}\partial_{i_1}\cdots\partial_{i_r}\mathcal{E}_{ij}
\end{equation}
with $c_r^{(\mathcal I)} = 1/(r+2)!$. For magnetic multipoles:
\begin{equation}
    S_{\rm p} \supset \frac{1}{2}\int dt\, c_r^{(J)}
    {\mathcal J}^{iji_1\cdots i_r}\partial_{i_1}\cdots\partial_{i_r}\mathcal{B}_{ij}
\end{equation}
with $c_r^{(\mathcal J)} = 2(r+2)/(r+3)!$.

\subsection{Feynman Rules}

\subsubsection*{Propagators}

\textbf{Radiative propagator} (for $\sigma_{ij}$ modes):
\begin{equation}
    G_R(\omega,\mathbf{k}) = \frac{i}{\mathbf{k}^2-(\omega+i0^+)^2}
\end{equation}

\textbf{Static propagator} (for energy insertion, zero frequency):
\begin{equation}
    G_{\rm static}(\mathbf{q}) = \frac i{\mathbf{q}^2}
\end{equation}

\subsubsection*{Source Vertices}

\textbf{Mass insertion:}
\begin{equation}
    V_M = -\frac{M}{\Lambda}
\end{equation}
This couples only to the $\phi$ mode and carries zero frequency.

\textbf{Electric quadrupole} ($r=0$, $c_0^{(I)} = 1/2$):
\begin{equation}
    V_{{\mathcal I}^{ij}} = \frac{-i}{2\Lambda}\cdot\frac{1}{2}\cdot\omega^2\cdot
    {\mathcal I}^{ij}(\omega)\cdot\sigma^*_{ij}
\end{equation}

\textbf{Electric octupole} ($r=1$, $c_1^{(\mathcal I)} = 1/6$):
\begin{equation}
    V_{{\mathcal I}^{ijk}} = \frac{-1}{2\Lambda}\cdot\frac{1}{6}\cdot\omega^2\cdot
    k_l\cdot {\mathcal I}^{ijl}(\omega)
\end{equation}
The extra $k_l$ comes from the spatial derivative $\partial_l$ in the
multipole action acting on the graviton field, bringing one power of
graviton momentum $k_l$ with $|\mathbf{k}|=\omega$ on-shell.

\textbf{Magnetic quadrupole} ($r=0$, $c_0^{(\mathcal J)} = 2/3$):
\begin{equation}
    V_{\mathcal J^{ij}} = \frac{-i}{2\Lambda}\cdot\frac{2}{3}\cdot\omega\cdot
    \epsilon_{iab}k_b\cdot {\mathcal J}^{ia}(\omega)
\end{equation}
The $\epsilon_{iab}k_b$ factor carries the \emph{external} graviton
momentum $k_b$ with $|\mathbf{k}| = \omega$ on-shell. This is because
the magnetic coupling in the action is $S_p\supset\int dt\,
{\mathcal J}^{ij}\mathcal{B}_{ij}$, and the spatial derivative acts on the
\emph{emitted graviton field}, bringing $k_b$ in momentum space.

\section{Master Integrals} \label{sec:app_master}

This appendix collects the two master integrals that appear in the main
computation: the one-loop mixed retarded-static integral $I_1(\omega)$,
which is the source of all IR divergences and hereditary logarithms, and the
off-shell retarded-retarded auxiliary bubble $J_{11}(\omega_1,\omega_2)$.
The latter must be combined with the full vertex numerator before taking the
external line on shell. Both are evaluated in $d = 3-\epsilon$ spatial
dimensions with $\int_\mathbf{q} =  \frac{d^d q}{(2\pi)^d}$.

The one-loop master integral with one retarded and one static propagator \cite{Foffa:2011np}:
\begin{equation}
    I_1(\omega) = \int_\mathbf{q}\frac{1}{(\mathbf{k}-\mathbf{q})^2
    [\mathbf{q}^2-(\omega+i0^+)^2]}
    \overset{k\to \omega}{=} \frac{[-(\omega+i0^+)^2]^{d/2-2}
    \Gamma(2-d/2)\Gamma(d-3)}{(4\pi)^{d/2}\Gamma(d-2)}
    \label{eq:master_integral_tail}
\end{equation}
At $d = 3-\epsilon$, evaluating each $\Gamma$ function\footnote{For $d=3$, the mixed massive-massless bubble develops a logarithmic singularity on shell,
$|\mathbf k|=|\omega|$, from the region $\mathbf q\to \mathbf k$, where the
massless and massive denominators vanish simultaneously. In dimensional
regularization this appears as a $1/\epsilon$ infrared-type pole after shifting
to the variable $\mathbf r=\mathbf k-\mathbf q$, so that the singular
region is $r\to 0$. Thus the pole is often denoted $1/\epsilon_{\rm IR}$,
although its physical origin is an on-shell pinch singularity.}:
\begin{align}
    [-(\omega+i0^+)^2]^{d/2-2} &= \frac{1}{\omega}
    \left[-\frac{(\omega+i0^+)^2}{\tilde\mu^2}\right]^{-\epsilon/2}\\
    \Gamma(2-d/2) &= \Gamma(1/2+\epsilon/2) \to \sqrt{\pi}\\
    \Gamma(d-3) &= \Gamma(-\epsilon) \to -\frac{1}{\epsilon_{\rm IR}}\\
    \Gamma(d-2) &= \Gamma(1-\epsilon) \to 1\\
    (4\pi)^{d/2} &\to 8\pi^{3/2}
\end{align}
Combining and expanding the scale factor:
\begin{equation}
    \left[-\frac{(\omega+i0^+)^2}{\tilde\mu^2}\right]^{-\epsilon/2}
    = 1 - \frac{\epsilon}{2}\log\frac{\omega^2}{\tilde\mu^2}
    + \frac{i\pi\epsilon}{2}{\rm sgn}(\omega) + O(\epsilon^2)
\end{equation}
So:
\begin{equation}
\boxed{
    I_1(\omega) = \frac{-1}{8\pi\omega}\left[\frac{1}{\epsilon_{\rm IR}}
    - \frac{1}{2}\log\frac{\omega^2}{\tilde\mu^2}
    + \frac{i\pi}{2}{\rm sgn}(\omega) + O(\epsilon)\right]
    }
    \label{eq:I1}
\end{equation}
This integral encodes the \textbf{single IR pole} from the static
propagator and the associated physical $\log\omega$.

\subsection*{Factorization at the outer tail pinch singularity}

For the all-tensor channel of the tail-of-memory graph, we define the
TT-projected memory form factor
\begin{equation}
 {\cal F}_{ij}(\omega,\mathbf p)
 \equiv \Pi_{ijkl}(\hat n)\,
 {\cal H}^{\rm mem}_{kl}(\omega,\mathbf p).
 \label{eq:app_tom_formfactor}
\end{equation}
The exact rearrangement of the two-loop graph in
\eqref{eq:tom_memory_factorization} reads
\begin{equation}
 \left.i{\cal A}_{\rm tom}^{\rm TT}\right|_{\text{IR}}
 =2i\,\frac{M\omega^2}{\Lambda^2}\,
 \sigma^*_{ij}\int_{\mathbf p}
 \frac{{\cal F}_{ij}(\omega,\mathbf p)}
 {[\mathbf p^2-(\omega+i0)^2](\mathbf k-\mathbf p)^2}.
 \label{eq:app_tom_factorization}
\end{equation}
Writing $\mathbf p=\mathbf k-\mathbf r$ and imposing
$\mathbf k^2=\omega^2$, the two denominators are
$(-2\mathbf k\cdot\mathbf r+\mathbf r^2)\mathbf r^2$.  Since the
fully assembled TT form factor is regular at this point,
\begin{equation}
 {\cal F}_{ij}(\omega,\mathbf k-\mathbf r)
 ={\cal F}_{ij}(\omega,\mathbf k)
 +r_a\,{\cal F}^{a}_{ij}(\omega,\mathbf k)+O(r^2).
 \label{eq:app_tom_formfactor_expansion}
\end{equation}
The constant term gives the master integral $I_1(\omega)$.  Each subsequent
term contains at least one power of $\mathbf r$ and is regular at the
leading pinch singularity; for example, the linear term scales radially as
\begin{equation}
 d^{3-\epsilon}r\,
 \frac{r}{\mathbf r^2(-2\mathbf k\cdot\mathbf r+\mathbf r^2)}
 \overset{r\to 0}{\sim} dr\,r^{-\epsilon}.
 \label{eq:app_tom_subleading_scaling}
\end{equation}
At the constant term, the on-shell relation
\begin{equation}
 \sigma^*_{ij}{\cal F}_{ij}(\omega,\mathbf k)
 =i{\cal A}_{\rm mem}^{\rm TT}(\omega,\hat n)
 \label{eq:app_tom_onshell_memory}
\end{equation}
therefore yields
\begin{equation}
 \left.i{\cal A}_{\rm tom}^{\rm TT}\right|_{\rm IR}
 =2i\,\frac{M\omega^2}{\Lambda^2}\,
 i{\cal A}_{\rm mem}^{\rm TT}(\omega,\hat n)\,I_1(\omega)
 +\text{terms subleading at the pinch singularity}.
 \label{eq:app_tom_I1}
\end{equation}
The on-shell limit in \eqref{eq:app_tom_onshell_memory} is taken only after
assembling the complete TT memory subgraph.  In particular, it cannot be
applied to an individual off-shell tensor-reduction coefficient such as
$B_0=J_{11}$.

Next, we consider the spatial loop integral involving two retarded propagators
carrying source frequencies $\omega_1=\omega'$ and
$\omega_2=\omega-\omega'$.
Away from its on-shell threshold, this auxiliary integral is UV and IR finite.
\begin{equation}
J_{11}(\omega_1,\omega_2)\equiv 
\int_{\mathbf q}
\frac{1}{
\big[\mathbf{q}^2-(\omega_1+i0^+)^2\big]
\big[(\mathbf k-\mathbf q)^2-(\omega_2+i0^+)^2\big]
}.
\end{equation}

We define
\begin{equation}
m_1\equiv \omega_1+i0^+, 
\qquad 
m_2\equiv \omega_2+i0^+,
\qquad
k=|\mathbf k|.
\end{equation}

\subsubsection*{Feynman parametrization}

Using
\begin{equation}
\frac{1}{AB}=\int_0^1 dx\,\frac{1}{[xA+(1-x)B]^2},
\end{equation}
we obtain
\begin{equation}
J_{11}
=
\int_0^1 dx
\int_{\mathbf q}
\frac{1}{\big[x(q^2-m_1^2)+(1-x)((\mathbf k-\mathbf q)^2-m_2^2)\big]^2}.
\end{equation}

Expanding $(\mathbf k-\mathbf q)^2=q^2-2\mathbf k\!\cdot\!\mathbf q+k^2$ and shifting
\begin{equation}
\boldsymbol\ell\equiv \mathbf q-(1-x)\mathbf k,
\end{equation}
the denominator becomes
\begin{equation}
xA+(1-x)B
=
\ell^2-\Delta(x),
\end{equation}
where
\begin{equation}
\Delta(x)=x m_1^2+(1-x)m_2^2-x(1-x)k^2 .
\end{equation}

Thus
\begin{equation}
J_{11}
=
\int_0^1 dx
\int_{\boldsymbol\ell}\frac{1}{(\ell^2-\Delta(x))^2}.
\end{equation}

\subsubsection*{General $d$ result}

Using the standard integral
\begin{equation}
\int \frac{d^d\ell}{(2\pi)^d}\frac{1}{(\ell^2+ \Delta)^\nu}
=
\frac{1}{(4\pi)^{d/2}}
\frac{\Gamma(\nu-d/2)}{\Gamma(\nu)}
\Delta^{d/2-\nu},
\end{equation}
with $\nu=2$, we obtain
\begin{equation} \label{eq:J11xeq}
\boxed{
J_{11}(\omega_1,\omega_2)
=
\frac{1}{(4\pi)^{d/2}}
\Gamma\!\left(2-\frac d2\right)
\int_0^1 dx\,
\Big[-
x m_1^2-(1-x)m_2^2+x(1-x)k^2
\Big]^{\frac d2-2}.
}
\end{equation}

To obtain a closed form, we write the quadratic polynomial as
\begin{equation}
-x m_1^2-(1-x)m_2^2+x(1-x)k^2
= - m_2^2(1-u x)(1-v x),
\end{equation}
where
\begin{equation}
u+v\equiv\frac{k^2+m_2^2-m_1^2}{m_2^2},
\qquad
uv \equiv \frac{k^2}{m_2^2}.
\end{equation}
Equivalently,
\begin{equation}
u,v \equiv
\frac{
k^2+m_2^2-m_1^2
\pm
\sqrt{\lambda}
}{
2m_2^2
},
\end{equation}
with
\begin{equation}
\lambda\equiv 
\big(k^2-(m_1+m_2)^2\big)\big(k^2-(m_1-m_2)^2\big).
\end{equation}

Hence
\begin{equation}
J_{11}
=
\frac{\Gamma\!\left(2-\frac d2\right)}{(4\pi)^{d/2}}
(-m_2^2)^{\frac d2-2}
\int_0^1 dx\,
(1-u x)^{\frac d2-2}(1-v x)^{\frac d2-2}.
\end{equation}

Using the Appell-$F_1$ representation
\begin{equation}
F_1(\alpha;\beta,\beta';\gamma;u,v)
=
\frac{\Gamma(\gamma)}{\Gamma(\alpha)\Gamma(\gamma-\alpha)}
\int_0^1 dt\,
t^{\alpha-1}(1-t)^{\gamma-\alpha-1}
(1-ut)^{-\beta}(1-vt)^{-\beta'},
\end{equation}
with
\begin{equation}
\alpha=1,
\qquad
\beta=\beta'=2-\frac d2,
\qquad
\gamma=2,
\end{equation}
we obtain the closed form
\begin{equation}
J_{11}(\omega_1,\omega_2)
=
\frac{\Gamma\!\left(2-\frac d2\right)}{(4\pi)^{d/2}}
\big[-(\omega_2+i0)^2\big]^{\frac d2-2}
F_1\!\left(
1;\,2-\frac d2,\,2-\frac d2;\,2;\,u,v
\right)\,.
\end{equation}

\paragraph{Special case $d=3$.}
For $d=3$, the Appell function collapses to the elementary logarithm
\begin{equation}
\boxed{
J_{11}(\omega_1,\omega_2)
=
\frac{1}{8\pi k}
\ln\!\left(
\frac{k+\omega_1+\omega_2}
{\omega_1+\omega_2-k}
\right),
}
\label{eq:mem_master}
\end{equation}
which can also be obtained by performing the $x$-integral of
\eqref{eq:J11xeq} for $d=3$. Eq.~(\ref{eq:mem_master}) is an off-shell expression whose value at threshold $k=\omega_1+\omega_2$ is logarithmically singular and therefore
must not be assigned a specific on-shell value.

\subsection*{Tadpoles in the memory tensor reduction}

The tensor reduction of the ordinary memory graph also involves the
one-propagator integral
\begin{equation}
T(\omega_a) \equiv \int_{\mathbf q}\frac{1}{\mathbf q^2-(\omega_a+i0^+)^2}
=\frac{\Gamma(1-d/2)}{(4\pi)^{d/2}}
\left[-(\omega_a+i0^+)^2\right]^{d/2-1}
=\frac{i\omega_a}{4\pi}+O(\epsilon).
\label{eq:tadpole_evaluation}
\end{equation}
Here dimensional regularization removes the power-divergent part; the
remaining term is finite and analytic in $\omega_a$.  Together with the
off-shell $J_{11}$ bubble above, these tadpoles enter the coefficients
$B_0$, $B_i$, and $B_{ij}$ in the memory tensor reduction of
section~\ref{sec:NRGR_memory}.  The cancellation of the $B_0=J_{11}$ terms in
the on-shell TT memory amplitude is a statement about that complete tensor
contraction, not an on-shell prescription for $J_{11}$ itself.

The two-longitudinal propagator loop integral \cite{Almeida:2021jyt}:
\begin{equation}
    J_{00}(k) \equiv \int_\mathbf{q}
    \frac{1}{\mathbf{q}^2
    (\mathbf{k}-\mathbf{q})^2}
    = \frac{(k^2)^{d/2-2}\Gamma(2-d/2)\Gamma^2(d/2-1)}
    {(4\pi)^{d/2}\Gamma(d-2)}\,,
\end{equation}
and $\lim_{d\to 3}J_{00}(\omega)=1/(8\omega)$.

\section{Polarization sectors and IR/UV structure} \label{sec:app_polarsector}

This appendix justifies the claim, used throughout the main text, that the
leading soft logarithms $\omega^{n-1}(\log\omega)^n$ of the
$n$-tail-of-memory amplitude arise exclusively from the all-tensor
$\{\sigma^2\}$ channel: the two internal retarded lines and the emitted
graviton are all $\sigma$ modes.  The remaining polarization sectors
($\{A\sigma\}$, $\{\phi\sigma\}$, $\{A\phi\}$, $\{A^2\}$) contribute UV
divergences and finite constants, but no additional leading soft logarithm.
The IR pole and the soft logarithm both originate from the master integral, $I_1(\omega)$.

\subsection*{Vertices in the tail-of-memory diagram:}

In the tail-of-memory diagram, there are two cubic bulk
vertices:
\begin{itemize}
  \item $V_1$ (momenta $\mathbf q$, $\mathbf p-\mathbf q$, $\mathbf p$): connects the two source multipole
        lines to the internal line $\mathbf p$.
  \item $V_2$ (momenta $\mathbf p$, $\mathbf k-\mathbf p$, $\mathbf k$): connects the internal line $\mathbf p$,
        the static $M$-insertion line $\mathbf k-\mathbf p$, and the external emitted
        graviton $\mathbf k$.
\end{itemize}
Two legs of $V_2$ are fixed by external conditions: the static line $k-p$
is forced to be $\phi$ (since $M$ couples exclusively to $h_{00} = -2\phi$),
and the external graviton $k$ is forced to be $\sigma$ (on-shell
transverse-traceless). The remaining internal line $p$ can be either
$\sigma$ or $A_i$, giving $V_2 = \sigma\sigma\phi$ or $V_2 = \sigma A\phi$
respectively, and fixing $V_1 = \sigma\sigma\sigma$ or $V_1 = \sigma\sigma A$
accordingly.

Following \cite{Almeida:2021jyt}, the full integrand decomposes according to
the polarizations of the two internal retarded lines ($\mathbf p$ and $\mathbf p-\mathbf q$).  We
denote the channel with these two lines and the emitted graviton all
$\sigma$ by $\{\sigma^2\}$; the other channels are denoted
$\{A\sigma\}$, $\{\phi\sigma\}$, $\{A\phi\}$, and $\{A^2\}$.

\subsection*{Why only $\{\sigma^2\}$ produces IR divergences:}

The IR singular region of the \emph{$p$-loop} is
$\mathbf r\equiv\mathbf k-\mathbf p\to0$: the static propagator is then
singular while the retarded propagator is on shell, $\mathbf p^2\to\omega^2$.
For this region to produce an IR divergence, the $p$-loop numerator must be free of
loop momentum after IBP; otherwise powers of $\mathbf r$ remove the singularity.

In the $\{\sigma^2\}$ channel:
\begin{itemize}
  \item The retarded $p$-line is a $\sigma$ propagator,
        $1/[\mathbf p^2-(\omega+i0^+)^2]$; the other retarded line entering
        the $q$ subloop carries the source frequency $\omega'$.
  \item The $\sigma\sigma\phi$ vertex $V_2$, after IBP, reduces to $\omega^2$
        --- a constant with respect to loop momentum $p$. So the $p$-loop
        integrand has \emph{no loop momentum in the numerator}, and the IR
        divergence from $\mathbf r\to0$ survives. 
  \item The complete TT memory form factor is regular at $\mathbf p=\mathbf
        k$.  Its constant term multiplies $I_1(\omega)$; terms proportional
        to $\mathbf k-\mathbf p$ are subleading at the pinch singularity $\mathbf{r} \rightarrow 0$, as shown in
        Appendix~\ref{sec:app_master}.
\end{itemize}
In contrast, for any sector involving other polarizations:
\begin{itemize}
  \item The $\sigma A\phi$ vertex at $V_2$ cannot be simplified to a
        constant via IBP — it necessarily retains loop momentum in the
        $p$-loop numerator (it requires a derivative to form a Lorentz scalar). The $\sigma\phi^2$ involves  either a trace of $\sigma$ or
        $\phi$ 3-momenta, thus killing the IR divergence.
  \item These extra powers of loop momentum remove the $\mathbf r\to0$
        singularity, producing no IR divergence.
\end{itemize}
This is also confirmed explicitly in \cite{Almeida:2021jyt}.

\subsection*{Why only $\{\sigma^2\}$ produces soft logarithms:}

The soft logarithms $\log\omega$ are \emph{inseparable} from the IR poles
--- they are the $\epsilon^0$ part of the same master integral $I_1(\omega)$ \eqref{eq:master_integral_tail}
that produces the IR pole.
Since the IR pole and the $\log\omega$ both come from the same integral
$I_1(\omega)$ in the $p$-loop, and this pinch survives only in the
$\{\sigma^2\}$ channel, the soft logarithms also come \emph{uniquely} from
$\{\sigma^2\}$ --- i.e.\ from $V_1 =
\sigma\sigma\sigma$ and $V_2 = \sigma\sigma\phi$.

The $\{A\sigma\}$ and other sectors cannot produce $\log\omega$ terms for
exactly the same reason they cannot produce IR poles: the extra loop
momentum in the p-loop numerator kills both the IR divergence and the
associated logarithm.

\subsection*{IR and UV structure at each order:}

The full amplitude at each order in $GM$ receives contributions from all
polarization sectors. The overall factor $[-(\omega+i0^+)^2/\tilde\mu^2]^\epsilon$
from dimensional regularization expands as:
\begin{equation}
  \left[-\frac{(\omega+i0^+)^2}{\tilde\mu^2}\right]^\epsilon
  = 1 + \epsilon\log\frac{\omega^2}{\tilde\mu^2}
  - \epsilon i\pi\sgn(\omega)+\frac{\epsilon^2}{2}\log^2\frac{\omega^2}{\tilde\mu^2} + \cdots
\end{equation}
and multiplies the pole structure. We now analyze each order:

\paragraph{Tail-of-memory ($n=1$, one $M$ insertion):}
There is no UV divergence at this order.  In the $\{\sigma^2\}$ channel, its
leading IR part is proportional to the canonical master-integral factor:
\begin{equation}
  iA_1\big|_{\rm IR}
  \;\propto\; iGM\,\sigma^{*ij}\mathcal{Q}_{ij}(\omega)
  \mathcal L(\omega).
\end{equation}
The remaining sectors contribute only finite terms at this order.  The
overall finite coefficient, including the PN--MPM constant, requires the full
vertex contraction.

\paragraph{Tail-of-tail-of-memory ($n=2$, two $M$ insertions):}
At $n=2$, the leading IR part of the $\{\sigma^2\}$ channel is proportional to
$[\mathcal L(\omega)]^2$, whereas the other polarization sectors can produce
a UV pole but no IR poles:
\begin{equation}
  [\mathcal L(\omega)]^2
  =\frac{1}{\epsilon_{\rm IR}^2}
  +\frac{2\alpha(\omega)}{\epsilon_{\rm IR}}
  +\alpha(\omega)^2,
  \qquad
  \alpha(\omega)\equiv-\frac 12\log\frac{\omega^2}{\tilde\mu^2}
  +\frac{i\pi}2\mathrm{sgn}(\omega).
\end{equation}
Accordingly, the full amplitude has the schematic separation
\begin{equation}
  iA_2\;\sim\;(iGM)^2\omega\,\sigma^{*ij}\mathcal{Q}_{ij}(\omega)
  \left[C_{2,{\rm IR}}\mathcal L^2(\omega)
  +\frac{\alpha_c^{(2)}}{\epsilon_{\rm UV}}+\Delta_2\right],
\end{equation}
where $C_{2,{\rm IR}}$ is fixed by the complete $\{\sigma^2\}$ contraction,
$\alpha_c^{(2)}$ is the UV coefficient, and $\Delta_2$ is finite.  The
$\log^2\omega$ term belongs to the IR factor, while the UV pole generates a RG logarithm after regularization.

\paragraph{3-tail-of-memory ($n=3$, three $M$ insertions):}
The same separation persists:
\begin{equation}
  iA_3\;\sim\;(iGM)^3\omega^2\,\sigma^{*ij}\mathcal Q_{ij}(\omega)
  \left[C_{3,{\rm IR}}\mathcal L^3(\omega)
  +\frac{\alpha_c^{(3)}}{\epsilon_{\rm UV}}+\Delta_3\right].
\end{equation}
The $\log^3\omega$ term is carried by the $\{\sigma^2\}$ IR factor; the UV
sector again produces a RG logarithm.

\subsection*{The general $n$-tail-of-memory:}

At leading IR order, the general amplitude has the form
\begin{equation}
  iA_n\big|_{\rm IR}
  \;\propto\;(iGM)^n\,\omega^{n-1}\,[\mathcal L(\omega)]^n
  \sigma^{*ij}\mathcal Q_{ij}(\omega).
\end{equation}
Each $\sigma\sigma\phi$ vertex supplies $\omega^2$ and each $I_1(\omega)$
supplies $1/\omega$; the remaining finite coefficient is fixed by the full
vertex contraction.  Writing $\mathcal L=1/\epsilon_{\rm IR}+\alpha$, the pole
structure is
\begin{equation}
  [\mathcal L]^n=\sum_{j=0}^n\binom{n}{j}
  \frac{\alpha(\omega)^j}{\epsilon_{\rm IR}^{n-j}}.
\end{equation}
\begin{itemize}
  \item The $n$-th IR pole $1/\epsilon_{\rm IR}^n$ and its logarithmic
        descendants arise in the $\{\sigma^2\}$ channel.
  \item All lower IR poles: fixed by exponentiation using results from
        lower orders $n'< n$.
  \item The single UV pole $\alpha_c^{(n)}/\epsilon_{UV}$: first appearing at
        each order $n\geq 2$, isolated by subtraction, renormalized into
        bare multipole.
  \item The leading soft log $\log^n\omega$ is fixed by the
        $\{\sigma^2\}$ IR factor.
  \item a UV log contribution, subleading with respect to the IR logs.
\end{itemize}
Thus the leading soft scaling is
\begin{equation}
  \boxed{iA_n\big|_{\omega\to 0} \sim (GM)^n\cdot\omega^{n-1}\cdot\log^n\omega}
\end{equation}
Within the leading-IR analysis, the $\{\sigma^2\}$ channel is the unique source
of the IR poles and the leading soft logarithms. The $\{A\sigma\}$ and
remaining sectors contribute UV poles at $n\geq2$ and finite remainders.

\end{appendix}

\bibliographystyle{JHEP}

\bibliography{references}

\end{document}